\documentclass[aps,pra,twocolumn,superscriptaddress]{revtex4-1}

\usepackage{graphicx,amsmath,hyperref, todonotes}

\newcommand{\psic}{\hat{\psi}^\dagger}
\newcommand{\psia}{\hat{\psi}^{}}

\newcommand{\bd}{\hat{b}^\dagger}

\newcommand{\hb}{\hat{b}^{}}

\newcommand{\swmat}{\hat{G}}

\renewcommand{\vec}[1]{\mathbf{#1}}

\usepackage{soul,color}

\begin{document}

\title{Excitonic spectral features in strongly-coupled organic polaritons}

\author{Justyna A. \'Cwik}
\affiliation{SUPA, School of Physics and Astronomy, University of St Andrews, St Andrews, KY16 9SS, United Kingdom}
\author{Peter Kirton}
\affiliation{SUPA, School of Physics and Astronomy, University of St Andrews, St Andrews, KY16 9SS, United Kingdom}
\author{Simone De Liberato}
\affiliation{School of Physics and Astronomy, University of Southampton, Southampton, SO17 1BJ, United Kingdom}
\author{Jonathan Keeling}
\affiliation{SUPA, School of Physics and Astronomy, University of St Andrews, St Andrews, KY16 9SS, United Kingdom}

\date{\today}

\begin{abstract}
  Starting from a microscopic model, we investigate the optical spectra of 
  molecules in strongly-coupled organic microcavities examining how they might self-consistently adapt their coupling to light. We consider both rotational and vibrational degrees of freedom, focusing on features which can be seen in the peak in the center of the spectrum at the bare excitonic frequency.  In
  both cases we find that the matter-light coupling can lead to a
  self-consistent change of the molecular states, with consequent
  temperature-dependent signatures in the absorption
  spectrum. However, for typical parameters, these effects are much
  too weak to explain recent measurements.  We show that another
  mechanism which naturally arises from our model of vibrationally dressed polaritons  has the right magnitude and temperature dependence to
  be at the origin of the observed data.
\end{abstract}
\maketitle

\section{Introduction}
\label{sec:introduction}

When matter is strongly coupled to light, the interaction cannot 
simply be thought of in terms of absorption and
emission processes. Instead we must consider the
eigenstates of the fully coupled matter-light system.  The
paradigmatic example of this is the existence of exciton polaritons,
hybrid matter-light particles formed by the strong interaction
between excitons and photons~\cite{hopfield58,pekar58}.  Matter-light
coupling can be engineered by confining light in optical cavities, so
as to modify the density of states and the coupling to matter.  For
weak coupling, or a bad cavity, cavity losses are fast so one can
eliminate virtual processes where photons are in the cavity. This gives
Fermi's golden rule, but with the cavity density of states modifying
the emission rate, as first discussed by Purcell~\cite{purcell46}.
When coupling is strong, first order perturbation theory (i.e.\ Fermi's
golden rule) fails, as there can instead be coherent emission and
reabsorption of photons before light leaks out of the
cavity~\cite{Bjork1991,weisbuch92}.

A natural context in which strong matter-light coupling arises is
between organic molecules and light in semiconductor microcavities.
Because of the existence of conjugated $\pi$ bonds in organic
molecules, electronic transitions can acquire large dipole
moments~\cite{AgranovichRussian,Davydov1971,Agranovich2009a}, leading
to very strong coupling to light.  When such molecules are placed in
optical microcavities this leads to huge polariton
splittings~\cite{lidzey98,lidzey99,Schwartz11,Tischler2005}.  These
scales allow such experiments to be performed at room temperature,
whereas for many inorganic materials, cryogenic temperatures are
required.  The polariton splitting is due to a collective phenomenon:
the electronic transitions of many molecules couple to radiation, and
as such the polariton splitting grows as the square root of the
molecule density.  In contrast, in weak coupling, the rate at which
one molecule emits is independent of whether or not any other
molecules are present~\cite{dicke54}.

Much of the recent work on organic microcavity polaritons (see, e.g.\
Ref.~\cite{Michetti2015} for a recent review) has been focused on
condensation and
lasing~\cite{Kena-Cohen2008,Forrest2010,Plumhof2013,Daskalakis2014},
involving a strongly pumped system, and the appearance of macroscopic
quantum coherence.  There has however also been significant recent
work on the effects of matter-light coupling in the vacuum state,
i.e.\ without strong pumping.  Such work aims to understand how the
physical and chemical properties of organic molecules are affected by
strong coupling to electromagnetic modes.  Examples of this include
modifying the rates of photochemical reactions~\cite{hutchison12}, or
modifying the transport properties of organic
semiconductors~\cite{Orgiu2014,Schachenmayer2014,Feist2014}.  More
recently, there has also been experimental~\cite{Shalabney2014} and
theoretical~\cite{Roelli2014,Pino2015, Pino2015b} work on coupling the
vibrational state of organic molecules to infra-red radiation, leading
to molecular optomechanics. Theoretical work~\cite{Spano2015} has also studied how
strong matter-light coupling to electronic states can suppress the
effects of disorder and vibronic features in the polariton spectrum.
Of particular interest for the present paper is a recent work from the
Ebbesen group~\cite{Canaguier-Durand2013}, in which the optical
spectra of strongly-coupled organic microcavities were studied by
varying molecular concentration and temperature, and paying particular
attention to the relative weights of the resonant features in the
absorption spectra: the two polariton peaks, and a third peak at the
bare excitonic energy~\cite{Schwartz2013a}. 

Our aim in this
manuscript is to examine the behavior of such strongly-coupled organic microcavities starting
from various microscopic models, allowing quantitative predictions of
the extent to which a self-consistent adaptation of the molecular
state, driven by coupling with light, may occur.


To understand the variation of the optical spectra with 
both concentration and temperature the models which we consider 
all contain a variable degree of
coupling to light.
This is because a molecule that has strictly zero coupling to light is
not visible in the absorption spectrum, while molecules with a small
but non-zero coupling will lead to absorption at the bare molecular
energy.  In order that the coupling to light can vary self
consistently (in response to the Rabi splitting), it must depend on
some adaptable feature of the state or environment of the molecule.
i.e.\ there must be some physical property that can vary, which
determines the strength of matter-light coupling. We refer to this
concept hereafter as ``self-consistent molecular adaptation''.
We refer to this process as ``self-consistent'' because the
effective matter-light coupling depends on (some aspect of) the
molecular state, and the molecular state is modified because of how
its energy depends on the matter-light coupling.
In the first part of this manuscript we
investigate in detail two candidates that could lead to
self-consistent adaptation: rotational and vibrational degrees of
freedom, we also consider an extension of these models to generic (classical)
aspects of the molecules physical or chemical state.  While we do
find a temperature dependence of the optical spectra, the involved
energy scales turn out to be incompatible with the observation of
Ref.~\cite{Canaguier-Durand2013}. In the final part of the present
work, we examine how our model of vibrationally dressed polaritons naturally predicts an effect
whose energy scale is of the right magnitude to explain the data. 
This effect does not involve the renormalization of the coupling strength, but instead involves the effect of vibrational replicas and their coupling to the excitonic transition on the optical spectra.  Our 
results thus show a first example of the rich, and presently poorly
understood behavior that can stem from the interplay of strong
matter-light coupling with strong coupling to
vibrational/conformational modes of the molecules.

We start by noting that the existence of a peak at the bare energy of the exciton, brought forward as evidence of novel physics in Ref.~\cite{Canaguier-Durand2013}, is not
unexpected; such a ``residual excitonic peak'' has been seen in many
cases, for example Ref.~\cite{houdre96} discussed theoretically, and
demonstrated experimentally, the appearance of such a feature in a
GaAs/AlGaAs heterostructure, containing quantum wells inside a DBR
microcavity.  While such a peak comes from the spectral weight of the exciton
line, it is important to note that this peak \emph{cannot} be viewed simply as
excitons which do not couple to light: if they did not couple,
they would not be visible in the absorption or transmission spectrum.
The origin of the peak can be understood physically
as coming from the subradiant excitonic states due to inhomogeneous broadening.  In a disordered
system, the coupling to the photon mode picks out a specific
superradiant state, which forms the polaritons, while the other
states --- orthogonal to this superradiant state --- remain at
the bare exciton energies.  However, because of energetic disorder, the
superradiant state is not an energy eigenstate, and a residual coupling 
between the superradiant and subradiant states exists, so that the 
spectral weight of the subradiant states is visible in
the optical spectrum~\cite{houdre96,Eastham2001,JKeeling2005}.

A simple analytical treatment shows that the weight of this residual
excitonic peak does decrease as the matter-light coupling increases.
The existence of this peak is thus consistent with the behavior seen
in~\citet{Canaguier-Durand2013}. However, on its own this explanation
cannot account for the temperature dependence observed, as the
residual excitonic peak should be unaffected by temperature, unless
$k_B T$ approaches optical energies, of the order of $1\text{eV}$ (for
comparison 300K$\approx25$meV).  One of the main goals of this paper
is to address how this temperature dependence may occur. 

As will become clear in the following, to discuss the microscopic
theory of such effects, it will be crucial to consider the physics of
ultrastrong matter-light coupling~\cite{Ciuti05, DeLiberato15}, and
the breakdown of the rotating wave approximation (RWA).  This requires
retaining ``counter-rotating'' terms in the matter-light coupling
Hamiltonian.  These terms, which involve simultaneous creation of
pairs of excitations, are typically considered to be non-resonant and
so are often neglected.  However, if the matter-light coupling is a
significant fraction of the bare exciton and photon energies, then
these terms have a non-negligible impact.  Such behavior has been seen
in both inorganic~\cite{Anappara09} and
organic~\cite{Schwartz11,Gambino14,Gubbin14} systems, with a current
record of a coupling strength $87\%$ of the bare oscillator
frequency~\cite{Maissen14}.  Our focus in this paper is on
the more typical regime where such counter-rotating terms cannot be 
neglected, but remain sufficiently small to be treated perturbatively.

The rest of the paper is structured in two main sections.  In
section~\ref{sec:rotational-freedom} we consider how temperature
dependence can arise due to self-consistent adaptation of the 
rotational and vibrational degrees of freedom of the molecules, via a
mechanism very similar to that proposed in
Ref.~\cite{Canaguier-Durand2013}.  For the orientational degree of
freedom, we consider both free molecules, and molecules with randomly
pinned orientations as appropriate in a polymer matrix. We will see
that in such systems we do predict a temperature dependence
of the residual excitonic peak.  However,
while this effect could potentially be observed in
other experimental realizations,
the energy scales (temperatures) required and the scaling with
molecular concentration are not compatible with the experimental
observations reported in Ref.~\cite{Canaguier-Durand2013}.

In section~\ref{sec:vibrational-freedom} we instead consider a
different effect, arising from the interplay of vibrational modes with
the matter-light coupling, which is able to reproduce similar behavior
to that observed in experiments.  Specifically we find that
vibrational excitations dress the residual excitonic peak in a
strongly temperature dependent manner.  Moreover, the form of the
vibrational dressed spectrum shows that the spectral feature at the
exciton energy can have a more complex interpretation than that
previously considered~\cite{houdre96}.

A brief but self-contained account of the main theoretical methods
used throughout this paper is given in the appendices.

\section{Self-consistent molecular adaptation due to strong coupling}
\label{sec:rotational-freedom}

In this section we consider whether self-consistent molecular adaptation  can
enhance matter-light coupling by renormalizing the bare matter-light coupling strength.  We consider models
in which the effective matter-light coupling strength of a given
molecule depends on the configuration of that molecule, such as its
orientation, or its vibrational state.  We then ask how this same
matter-light coupling modifies the energy landscape for the auxiliary
parameters describing the configuration. This leads to the idea of
self consistency --- if strong coupling leads to a reduction of the ground state
energy, the energy landscape is deformed so as to favor auxiliary
parameters for which the effective matter-light coupling is as large
as possible.  Our aim is to derive this from a microscopic model, and
so quantify this effect.  In the following we consider two potential
scenarios involving adaptation of either orientational or
vibrational degrees of freedom.

If such a self-consistent enhancement of matter-light coupling occurs,
then this can lead to a temperature dependent effective
coupling, and thus to a temperature dependence of the residual excitonic
peak.  We show that such an effect exists, but that its strength is
relatively weak, and that the relevant energy scale shows no
collective enhancement, i.e.\ the presence of $N_m$ molecules does
not lead to a $N_m$ enhancement of this energy scale, because it must
compete with the extensive entropy gain from orientational or
vibrational disorder.  As such, while increasing the molecular
concentration will increase the polariton splitting, it has little
effect on the self-consistent orientation.  Changing the bare oscillator
strength of the molecules does however affect both the polariton
splitting and the self-consistent molecular adaptation energy scale.

Before introducing any auxiliary variables, the basic Hamiltonian
which we consider is an extended Dicke model, including diamagnetic
terms:
\begin{multline}
  \label{eq:1}
  \hat{H} = \sum_{\vec{k}} \omega_{\vec{k}} \psic_{\vec{k}} \psia_{\vec{k}} 
  + \sum_{n} \Biggl[
  \frac{\epsilon_n}{2} \sigma_n^z 
  \\+
  \sum_{\vec{k}}  g_{\vec{k},n} \varphi_{\vec{k}}(\vec{r}_n) \sigma^x_n
  +
  \left( \sum_{\vec{k}}
    \sqrt{D_{\vec{k}}} \varphi_{\vec{k}}(\vec{r}_n)
  \right)^2\Biggr],
\end{multline}
where the field $ \varphi_{\vec{k}}(\vec{r}) = \psic_{\vec{k}} e^{-i
  \vec{k}\cdot\vec{r}} + \psia_{\vec{k}}e^{i \vec{k}\cdot\vec{r}}$ is
written in terms of the bosonic creation and annihilation operators
$\psia_{\vec{k}},\psic_{\vec{k}}$ describing photon modes labeled by
their in-plane momentum $\vec{k}$, and energy $\omega_{\vec{k}}$. The Pauli matrices
$\sigma^{i=x,y,z}_n$ describe the electronic state of the molecules.
For completeness, we therefore also included the
diamagnetic terms, arising from the $\mathbf{A}^2$ term of the minimal coupling Hamiltonian~\cite{DeLiberato14}.

Furthermore, in order to consider varying the cavity mode volume while respecting the
Thomas-Reiche-Kuhn sum rule~\cite{Rzazewski1975} we use $g_{\vec{k},n}
= \sqrt{\epsilon_n D_{\vec{k}} f_n}$ where $0<f_n<1$ is the oscillator
strength of the given molecule.  The coefficient $D_{\vec k}$ depends
on the electric field strength of a single photon, and the properties
of the effective charges that respond to the field.  

If we assume an uniform distribution of molecules, momentum
conservation can be used to write the diamagnetic term in
Eq.~(\ref{eq:1}) as $N_m \sum_k D_{\vec{k}} (\psic_{\vec{k}} +
\psia_{-\vec{k}}) (\psic_{-\vec{k}} + \psia_{\vec{k}})$.  This term
can then be removed by a Bogoliubov transformation $\psia_{\vec{k}}
\to \cosh \theta_{\vec{k}} \psia_{\vec{k}} + \sinh \theta_{\vec{k}}
\psic_{\vec{k}}$, yielding the effective Hamiltonian:
\begin{equation}
  \label{eq:2}
    \hat{H} = \sum_{\vec{k}} \tilde{\omega}_{\vec{k}} \psic_{\vec{k}} \psia_{\vec{k}} 
  + \sum_{n} \hat{h}_n,
\end{equation}
where the on-site Hamiltonian is given by
\begin{equation}
	\hat{h}_n = \frac{\epsilon_n}{2} \sigma_n^z + 
    \sum_{\vec{k}}\tilde{g}_{\vec{k},n} \varphi(\vec{r}_n) \sigma^x_n,
\end{equation}
and the renormalized parameters $\tilde{\omega}_{\vec{k}},
\tilde{g}_{\vec{k}}$ are:
\begin{equation}
  \label{eq:14}
  \tilde{\omega}_{\vec{k}} =
  \sqrt{\omega_{\vec{k}}(\omega_{\vec{k}} + 4 N_m D_{\vec{k}})},
  \qquad
  \tilde{g}_{\vec{k},n} =
  g_{\vec{k},n}\sqrt{\frac{\omega_k}{\tilde{\omega}_k}},
\end{equation}
 with $N_m$ the number of molecules in the mode volume.

In the dipole approximation, for a molecule with a single mobile
electron $D_{\vec{k}} = \zeta / \omega_\vec{k} V$ with $V$ the mode volume
and $\zeta= \hbar^2 e^2 / (4 m_r \varepsilon_0)$
 quantifying the electronic
response of a single electron in terms of the vacuum permittivity
$\varepsilon_0$ and its reduced mass $m_r$. For molecules involving
many conjugated bonds, the coefficient $\zeta$ is replaced by a sum
over all mobile charges.  It is important to note that changing the
cavity length changes both the mode volume $V$ (and hence both
$D_{\vec k}$ and $g_{\vec{k},n}$) and the spectrum of photon modes
$\omega_\vec{k}$.  In the following numerical results we will fix the values
of the polariton splitting $g_{\vec{k},n}\sqrt{N_m}$, exciton energy
$\epsilon_n$, and coupling strength $f$, and use these to determine
$D_{\vec{k}}N_m$.

Before considering the role of vibrational and orientational degrees
of freedom, we consider how the existence of a temperature dependent
matter-light coupling strength, $\tilde{g}_{\text{eff},\vec{k},n}$
would be seen in the absorption spectrum. For this purpose, it is
sufficient to consider the absorption spectrum of Eq.~(\ref{eq:2}),
and its dependence on $\tilde{g}_{\vec{k},n}$.
Appendix~\ref{sec:absorpt-transm-refl} summarizes how the absorption
spectrum can be calculated, including the counter-rotating terms in
the matter-light coupling~\cite{Ciuti2006}. In performing these 
calculations, as noted above, it is necessary to include disorder 
in order to see the residual excitonic feature. We will consider 
disorder in the exciton energies, denoted by the energy distribution 
$h(\epsilon)$ For simplicity we consider only disorder in the energies
and we ignore the subleading effect of the exciton energy
distribution on the coupling strength $g_{\vec{k},n}$, by using
the averaged value $g_\vec{k}^2 \equiv \sum_n g_{\vec{k},n}^2/ N_m$.
 As discussed in the
appendix, we consider the quantity $a_{\vec{k}}(\nu) = -2
\Im[G^R_{\vec{k},xx}(\nu)]$, in terms of the retarded Green's
$G^R_{\vec{k},xx}(\nu)$, which is proportional to the absorption
spectrum for a good cavity.  The Green's function has the form:
\begin{equation}
  \label{eq:3}
  {G}^R_{\vec{k},xx}(\nu)  =
  \frac{2 {\omega}_k}{%
    \nu^2 - \tilde{\omega}_k^2 
    + 2{\omega}_k \Sigma_{\vec{k},xx}(\nu)    
    + i\tilde{\omega}_k \tilde{\kappa}(\nu)},
\end{equation}
where the excitonic self energy for Eq.~(\ref{eq:2}), can be written
as ${\Sigma}_{\vec{k},xx}={\Sigma}^{\text{RWA}}_{\vec{k},+-}(\nu) +
{\Sigma}^{\text{RWA}}_{\vec{k},+-}(-\nu)^\ast$ with:
\begin{equation}
  \label{eq:4}
  {\Sigma}^{\text{RWA}}_{\vec{k},+-}(\nu) =-  g^2_{\vec{k}}  N_m
  \int_{-\infty}^{\infty} \!\!\!\! d\epsilon h(\epsilon)
  \frac{\tanh(\beta \epsilon/2)}{%
    \nu+i\gamma^+ - \epsilon}.
\end{equation}
Here $\beta=(k_BT)^{-1}$ is the inverse temperature and $\gamma$ is the 
homogeneous linewidth of the excitons. In the following we will present 
results both for $\gamma=0^+$ and for small but non-zero $\gamma$ as indicated 
in the figure captions.

\begin{figure}[htpb]
  \centering
  \includegraphics[width=3.2in]{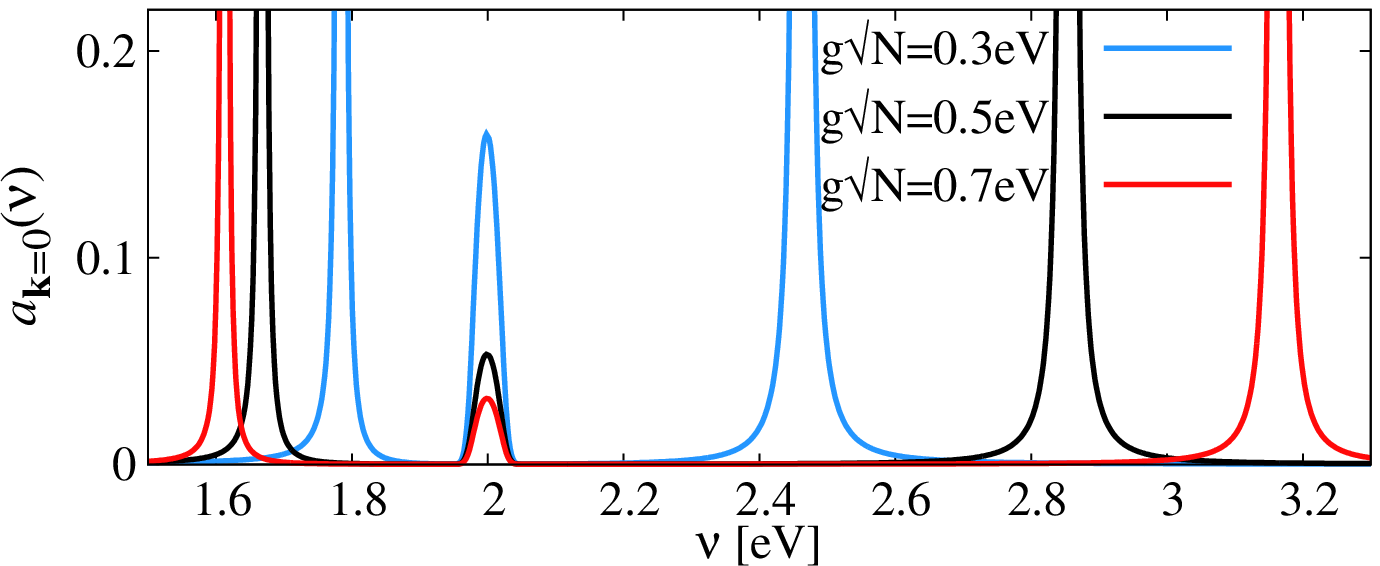}
  \includegraphics[width=3.2in]{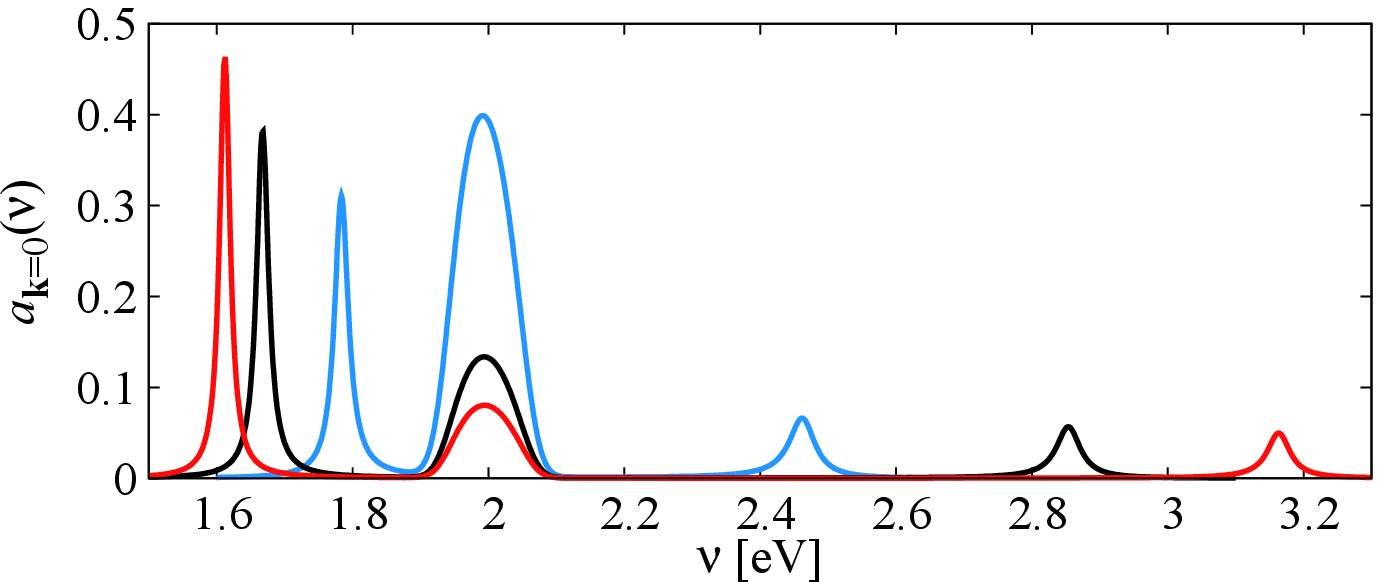}
  \caption{(color online) Evolution of $\vec{k}=0$ absorption spectrum, $a(\nu)$ vs
    polariton splitting $g_{\vec{k}=0} \sqrt{N_m}$ for the system with
    no auxiliary degrees of freedom.  Plotted for $k_B T=0.025$eV
    (i.e.\ $T=290$K), $f=0.05,\omega_{\vec{k}=0}=2.1$eV, and a truncated
    Gaussian distribution $h(\epsilon) \propto
    \Theta(\epsilon)e^{(\epsilon - \epsilon_0)^2/2\sigma^2}$ with
    $\epsilon_0=2.0$eV, $\sigma=0.01$eV. Panel (a) shows the results with a perfect cavity while (b) includes the effects of cavity losses at rate $\kappa=0.075$eV and excitonic non-radiative decay at rate $\gamma=10^{-4}$eV also the width of the energy distribution is larger, $\sigma=0.025$eV  as discussed in Appendix~\ref{sec:absorpt-transm-refl}.}
  \label{fig:abs_simple}
\end{figure}

The spectrum is shown in Fig.~\ref{fig:abs_simple}, focusing on the
residual excitonic peak, to show its dependence on the polariton
splitting.  As noted above, such a feature has been observed and
commented on several times before,
e.g.~\cite{houdre96,Eastham2001,JKeeling2005}.  When the polariton
splitting $g\sqrt{N_m}$ is increased, the splitting between the lower
and upper polaritons increases, and the exciton spectral weight of the
feature at the exciton energy decreases.  The asymmetry between the
shift of the lower and upper polaritons arises due to the diamagnetic
term $D_k$ renormalizing the photon energy.  If the coupling to light
is weak, so that the RWA is valid, this asymmetry vanishes. In Fig.~\ref{fig:abs_simple} (b) we show how the spectrum is modified by including the effects of cavity losses and non-radiative excitonic decay as described in Appendix~\ref{sec:absorpt-transm-refl}. The main effect these processes have is to broaden and thus reduce the height of the polariton peaks, there is also additional broadening to the central peak.

Because the only temperature dependence of Eq.~(\ref{eq:3},\ref{eq:4})
is via the combination $\tanh(\beta \epsilon/2)$, the spectrum is
temperature independent while $k_B T \ll \epsilon$ (with $\epsilon$ of the order of the exciton bare energy).  Therefore, as
noted in the introduction, the experimentally observed temperature
dependence cannot occur from this mechanism alone, unless the
effective value of $g_{\vec{k},n}$ is made temperature dependent via
its dependence on configuration. We now go on to consider if this effect is plausible.

\begin{figure}[htpb]
  \centering
  \includegraphics[width=3.2in]{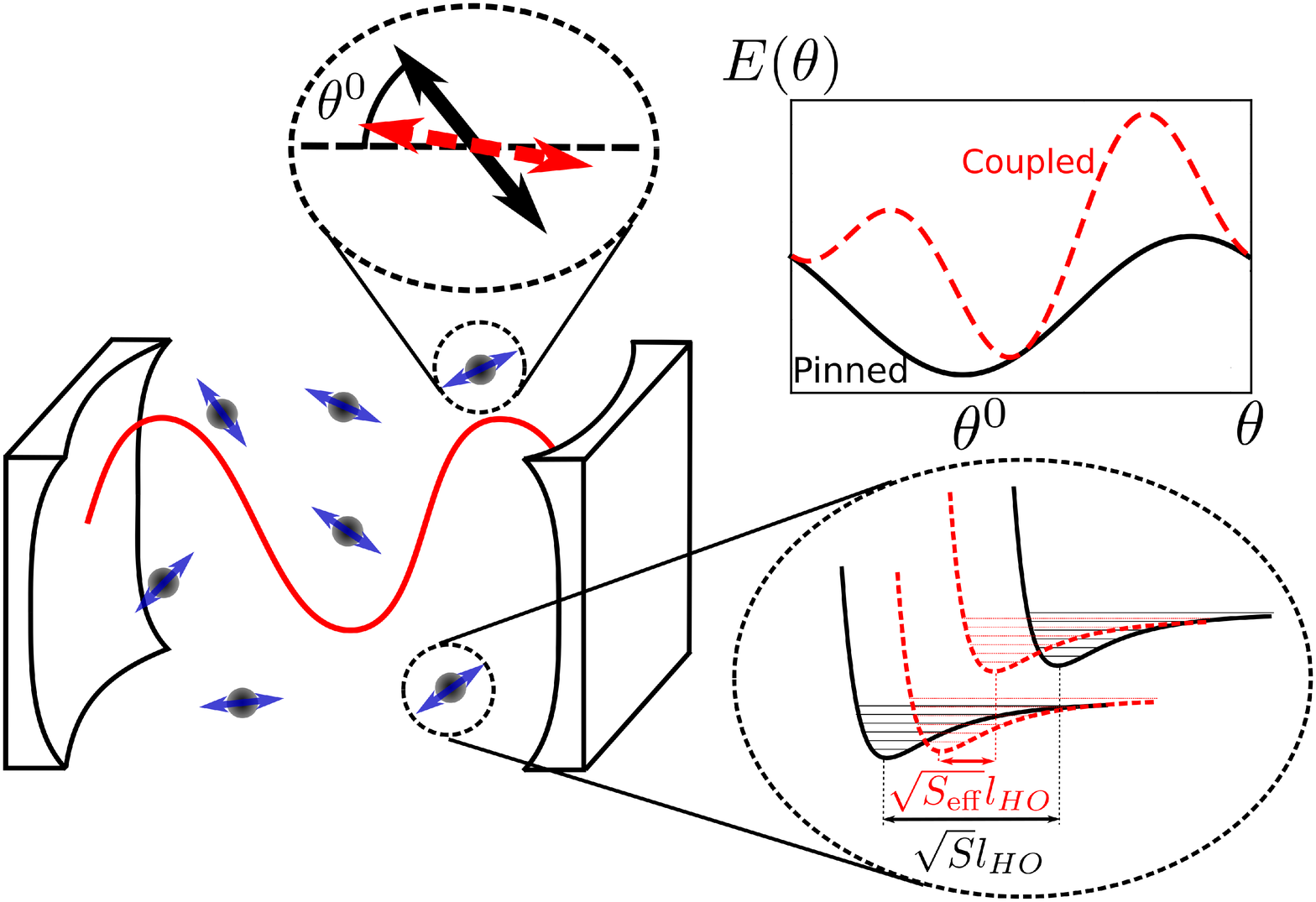}
  \caption{(color online) Schematic diagram showing orientational and vibrational degrees of freedom. The top right inset  shows the potential energy landscape with (dashed red line) and without (solid black line) a cavity. The other insets show (top) the orientational degree of freedom and (bottom-right) the vibrational degree of freedom illustrating how the coupling to light decreases the displacement between the ground and excited manifolds.  Shifts due to coupling to the cavity are exaggerated for clarity in all three insets.}
  \label{fig:cartoon}
\end{figure}

To consider molecular adaptation, the Hamiltonian in
Eq.~(\ref{eq:2}) will be modified to include either orientational or
vibrational degrees of freedom. These are illustrated in Fig.\ \ref{fig:cartoon}. We next introduce these modifications
and then discuss how they may be treated.

The Hamiltonian including an orientational degree of freedom takes the same form as Eq.~\eqref{eq:2} with the on-site part
\begin{align}
  \label{eq:6}
  \hat{h}_n &= 
  \frac{\epsilon_n}{2} \sigma_n^z 
  + 
  \Lambda_n(\theta_n)
  +
  \sum_{\vec{k}}
  \tilde{g}_{\vec{k},n} \cos(\theta_n) \varphi_{\vec{k}}(\vec{r}_n) \sigma^x_n,
\end{align}
where the angle $\theta_n$ parametrizes the orientation of the dipole moment of the $n$th
molecule with respect to the polarization of the cavity electric
field, thus reducing the oscillator strength.  The term
$\Lambda_n(\theta_n)$ represents the bare dependence of the
Hamiltonian on orientation. This term allows one to model pinning of the orientation $\theta_n$.
For simplicity we consider only a classical orientational degree of
freedom;  the corresponding quantum theory would require us to also
include a rotational kinetic energy term, and diagonalize the resulting
Hamiltonian. The effective matter-light coupling
strength, which depends on the distribution of angles
$\theta_n$ adopted by the molecule,  can be written as $\tilde{g}^2_{\vec{k},n,\text{eff}} =
\tilde{g}^2_{\vec{k},n} \langle\langle \cos^2(\theta_n)
\rangle\rangle$, where the double angle brackets represent both an ensemble and thermal
average.

To consider vibrational degrees of freedom, we again start from the
transformed Hamiltonian, Eq.~(\ref{eq:2}), and now consider
the following modification:
\begin{multline}
  \label{eq:7}
  \hat{h}_n =  
    \frac{\epsilon_n}{2} \sigma_n^z + 
    \sum_{\vec{k}} \tilde{g}_{\vec{k},n}  \varphi_\vec{k}(\vec{r}_n) \sigma^x_n +
    \\
    \sum_m \Omega_m \left( 
      \hat{b}^\dagger_{n,m} \hat{b}^{}_{n,m}
      +
      \frac{\sqrt{S_m}}{2} (\hat{b}^\dagger_{n,m} + \hat{b}^{}_{n,m}) \sigma^z_n
    \right).
\end{multline}
Here $\hat{b}_{n,m}, \hat{b}^\dagger_{n,m}$ describe the $m$th
harmonic vibrational mode of molecule $n$, with the mode having
frequency $\Omega_m$ and its coupling to the electronic state being
parametrized by the Huang-Rhys parameter $\sqrt{S_m}$.  In this case,
defining the effective oscillator strength is more involved: the
effective oscillator strength depends on the matrix element describing
the overlap between the vibrational states in the ground and excited
state manifold.  We return to this point in later sections.

For both the orientational and vibrational degrees of freedom, our aim
is to find how the matter-light coupling is self-consistently modified
by these auxiliary degrees of freedom: i.e.\ how the presence of
matter-light coupling modifies the distribution of orientational or
vibrational states, and how this in turn affects the effective
matter-light coupling strength. We consider a case without any strong
pumping, and with a temperature such that $k_B T \ll \omega,\epsilon$,
which is typically satisfied even at room temperature for organic
polaritons.  As such, the origin of the ``self-consistent'' dependence
of configuration on the matter-light coupling coupling arises due to
the existence of the counter-rotating terms in the original
Hamiltonian: If these terms were neglected then the energy in the
ground state sector can be trivially found as the ground state would correspond to
the empty state, and its energy would therefore not involve the matter-light
coupling strength at all~\cite{Ciuti05}.  The presence of counter-rotating terms
means that the ground state sector also involves an admixture of all
even parity sectors, and the degree of admixture depends on the
effective matter-light coupling.

As discussed below, while exact solutions are possible in some limiting
cases of the orientational problem, these are not generally possible
at finite temperature, nor for the vibrational problem.  This is
because thermal or quantum fluctuations of the auxiliary degrees of
freedom break translational symmetry, preventing simple exact
diagonalization.  As such, we proceed using the Schrieffer-Wolff
formalism~\cite{schrieffer66}, which allows us to consider perturbatively the effects of
these counter-rotating terms, and how they modify the energy landscape
seen by the auxiliary orientational or vibrational degrees of freedom.
For completeness, appendix~\ref{sec:schr-wolff-transf} provides a
brief summary of the Schrieffer-Wolff formalism.  The essential point
is to separate $\hat{H}=\hat{H}_0 + \hat{H}_1$, where $\hat{H}_1$ are
the terms treated perturbatively. At leading order this gives an
effective Hamiltonian:
\begin{equation}
  \label{eq:8}
  \hat{\tilde{H}}  \approx 
  \hat{H}_0 + \frac{i}{2}\left[\swmat,\hat{H}_1 \right],
  \quad
  \swmat: \ 
  [\swmat, \hat{H}_0] \equiv i \hat{H}_1.
\end{equation}
Taking the counter-rotating terms as $\hat{H}_1$, the perturbation
theory is controlled by the small parameter
$\tilde{g}_{\vec{k},n}/(\omega_{\vec{k}}+\epsilon_n)$, which is indeed
small for the physical parameters we consider~\footnote{Note however
  that the collective splitting $g_\vec{k}\sqrt{N_m}$ can 
  still be comparable to $\omega_k+\epsilon$, as is indeed the case
  for the parameters we consider.}.  We are thus in a regime
where the counter-rotating terms cannot be ignored, but where they can be
included perturbatively.  In the following sections we apply this
approach in turn to the orientational and vibrational degrees of freedom,
and see how the effective matter-light coupling can be derived
self-consistently.

\subsection{Orientational degrees of freedom}
\label{sec:orient-degr-freed}

As discussed above, we consider first the classical orientational
degrees of freedom $\theta_n$, subject to a pinning potential
$\Lambda_n(\theta_n)$. If all molecules are identical,
$\Lambda_n(\theta_n)=\Lambda(\theta_n)$, then one can find the zero
temperature ground state by choosing $\theta_n=\theta$. For the ground
state, all that is required is to find the quantum ground state of
Eq.~(\ref{eq:6}) as a function of $\theta$ and then minimize over
$\theta$.  Since Eq.~(\ref{eq:6}) is translationally invariant in the case
$\theta_n=\theta$, it is possible to find the exact ground state in the
bosonic approximation by Fourier transforming.  The bosonic
approximation assumes the occupation of each excited molecule is
small, a result that is valid unless $g_{\vec{k}} \gg
\omega_{\vec{k}},\epsilon$~\cite{Casanova10,DeLiberato14}.  However,
at finite temperature, even for identical molecules, it is crucial to
allow independent fluctuations of each $\theta_n$; assuming
$\theta_n=\theta$ massively underestimates the entropy at finite
temperature. 
At zero temperature, one may compare the exact solution to the
Schrieffer-Wolff perturbative expansion used below, and one  finds that
these indeed match to leading order.

For this rotational case, the form of $\swmat$ required in
Eq.~(\ref{eq:8}) can be found trivially, and the resulting Hamiltonian
can most conveniently be written as $\hat{\tilde{H}} = \sum_{\vec
  k}\tilde{\omega}_{\vec k} \psic_{\vec{k}} \psia_{\vec{k}} + \sum_n
\hat{\tilde{h}}_{n}$, with the molecular Hamiltonian in the form:
\begin{displaymath}
  \hat{\tilde{h}}_{n} = {\hat{h}}^{(0,\text{RWA})}_n
  - 
  \sum_{\vec{k}} 
  \frac{\tilde{g}_{\vec{k},n}^2 \cos^2(\theta_n)}{%
    \tilde{\omega}_{\vec{k}} + \epsilon_n},
\end{displaymath}
where ${\hat{h}}^{(0,\text{RWA})}_n$ is the bare molecular Hamiltonian,
including the RWA coupling to light
and including the pinning term $\Lambda_n(\theta_n)$.

In order to consider the thermal distribution of $\theta_n$, we must
specify the orientational potential $\Lambda_n(\theta_n)$.  We
consider a form
\begin{equation}
	\Lambda_n(\theta_n)=
        \lambda\sin^2\left(\frac{\theta_n-\theta_n^0}2\right),
\end{equation}
which tries to pin the molecules at angle $\theta_n^0$, relative to the cavity electric field, with strength
$\lambda$.  We may thus consider both the free orientation case,
$\lambda=0$, and the pinned case simultaneously. In what follows we will 
assume that the pinning angles have a uniform distribution such as would be found in a polymer matrix. This treatment, however, ignores effects which would be important in systems such as organic crystals in which the pinning angle has a fixed direction. The
energy landscape for each angle $\theta_n$, given the pinning angle
$\theta_n^0$ is then:
\begin{equation}
	E(\theta_n\vert\theta_n^0)=\lambda\sin^2\left(\frac{\theta_n-\theta_n^0}2\right)
	-\sum_{\vec{k}} 
        \frac{\tilde{g}_{\vec{k}}^2 \cos^2(\theta_n)}{%
          \tilde{\omega}_{\vec{k}} + \epsilon}.
\end{equation}
where for simplicity we have assumed that each molecule has the same values 
of $\epsilon, \tilde g_\vec{k}$ and the only disorder is in the
pinning angles. In the following we will define the quantity:
\begin{equation} \label{eq:defK0}
  K_0 \equiv
  \sum_{\vec{k}} 
  \frac{\tilde{g}_{\vec{k}}^2 }{(\tilde{\omega}_{\vec{k}} + \epsilon)}.
\end{equation}
The quantity $K_0$ characterizes the self-consistent energy favoring
alignment of molecules.  Replacing the summation by an integral, and
inserting the explicit forms of $g_{\vec k}$ and $D_{\vec k}$ written
above we have that
\begin{equation}
  \label{eq:intK0}
  K_0
  = 
  \frac{f \epsilon \zeta}{l_c}\int_0^{\Lambda}
  \frac{k d k}{(2\pi)} \frac{1}{\tilde{\omega}_k(\tilde{\omega}_{k} + \epsilon)},
\end{equation}
where $l_c$ is again the cavity length, $\zeta$ the combination
defined following Eq.~(\ref{eq:2}), and $\Lambda=2\pi/a_{\text{Bohr}}$
is a cutoff reflecting the breakdown of the dipole approximation.  To
evaluate such integrals it is useful to write the
dispersion in the form $\tilde{\omega}_k^2 = (\omega_0^2+c^2k^2) + 4
\zeta N_m/V$ which allows us to find the exact result
\begin{equation}
	K_0 = \frac{f\epsilon\zeta}{2\pi l_c c^2}\ln\left(\frac{\epsilon+\tilde\omega_{\Lambda}}{\epsilon+\tilde\omega_0}\right). \label{eq:K0exact}
\end{equation}

A notable feature of Eq.~(\ref{eq:K0exact}), as anticipated above, is
that this quantity \emph{does not simply increase} as one increases
the polariton splitting by varying the density of emitters, $N_m/V$.
There is a sub-leading dependence on the molecule density, via the
renormalization $\omega_{\vec{k}} \to \tilde{\omega}_{\vec{k}}$ given
in Eq.~(\ref{eq:14}).  However this effect is only significant in the
deep strong coupling limit~\cite{Casanova10,DeLiberato14}.  Physically
this lack of scaling with $N_m$ is because this ``molecular adaptation energy'' depends on the shift
seen for \emph{each} molecule.  Inserting typical experimental values for an organic system
$N_m/V=4.2\cdot 10^{25}\text{m}^{-3}$, $\epsilon=\omega_0=2.1$eV,
$l_c=145$nm, $f=0.5$, $a_{\text{Bohr}} = 2$nm, $\zeta=4.3$eV$^2$nm$^3$ which correspond to the
values extracted from Ref.~\cite{Canaguier-Durand2013}, one finds that
\begin{displaymath}
  K_0 = 6.3\cdot 10^{-4} \text{meV},
\end{displaymath}
which is much smaller than $k_BT$ at room temperature. 

As noted earlier, the effective matter-light coupling strength is
given by
\begin{math}
  {g_{\text{eff},\vec{k}}^2}={g^2_{\vec{k}}} 
  \langle\langle\cos^2(\theta_n) \rangle\rangle.
\end{math}
At zero temperature, this corresponds to minimizing the
energy, leading to $\lambda_n\sin(\theta_n - \theta_n^0) = - 2K_0
\sin(2\theta_n)$.  However, since the value of $K_0$ given above is
such that $K_0\ll k_B T$, it is crucial to consider finite
temperatures.  The smallness of the ratio $K_0/k_B T$ will also allow
us to make further perturbative expansions in the following.

Defining $ {g^2_{\text{eff}}}/{g^2} \equiv \langle\langle \cos^2(\theta_n)
  \rangle\rangle$, this ratio can be calculated as
\begin{equation}
  \frac{g_{\text{eff}}^2}{g^2}= 
  \frac{1}{2\pi\beta}
  \frac{d}{dK_0}\int d\theta^0
  \ln[\mathcal{Z}(\lambda, K_0,\theta^0)],
\end{equation}
where the partition function is
\begin{displaymath}
  \mathcal{Z}(\lambda, K_0, \theta^0)=\!\!
  \int\!\! d\theta \exp\left[
    \beta K_0\cos^2\theta-\beta\lambda\sin^2\left(\frac{\theta-\theta^0}2\right)
  \right].
\end{displaymath}
As noted above $\beta K_0\ll1$, and so we may Taylor expand in this
small parameter to get a closed form for $g_{\text{eff}}$. Assuming
that the pinning angle distribution $\chi(\theta^0)=1/2\pi$ is uniform
we obtain the simple result for the coupling
\begin{equation}
  \frac{g_{\text{eff}}^2}{g^2}= \frac{1}{2}\left[
    1+\frac{\beta K_0}{4}
    \left(
      1-\frac{I_2(\beta\lambda/2)^2}{I_0(\beta\lambda/2)^2}
    \right)
  \right]
\end{equation}
with $I_n(z)$ the imaginary Bessel function
\begin{equation}
  I_n(z) = \!\!\int\!\! d\theta \cos(n \theta) e^{z \cos \theta}.
\end{equation}
At this point we have made no assumption about the pinning strength
$\lambda$, and the value of the effective coupling is controlled by
the combination $\beta \lambda$. If the pinning is strong
$\beta\lambda\gg1$ then we find the asymptotic form
\begin{equation}
  \label{eq:9}
  \frac{g_{\text{eff}}^2}{g^2}=\frac{1}{2}\left(1+\frac{2K_0}{\lambda}\right).
\end{equation}
This no longer depends on temperature as the strong pinning limit
means entropy become unimportant.  Thus, in the $\lambda\to\infty$
limit the coupling takes on the isotropic value of $1/2$,
corresponding to the uniform distribution of angles $\theta_n =
\theta_n^0$.  The effective coupling increases as the pinning
$\lambda$ decreases.  In the limit of vanishing pinning $\lambda\to 0$,
the imaginary Bessel function $I_2(0)$ vanishes and so we have:
\begin{equation}
  \label{eq:10}
  \frac{g_{\text{eff}}^2}{g^2}=\frac{1}{2}\left(1+\frac{\beta K_0}{4}\right).
\end{equation}
Both Eq.~(\ref{eq:9}) and Eq.~(\ref{eq:10}) indicate that as long as
$\beta K_0 \ll 1$, the modification and temperature dependence of
$g_{\text{eff}}^2$ is very small. We note that while in the organic systems which we focus on here $K_0$ is relatively small, in systems which have 
very small mode volumes
 this parameter could be engineered to be much larger and hence the coupling strength renormalization would be much more pronounced.  i.e., since
there is no scaling with $N_m$, the crucial feature to see a strong renormalization is to minimize the mode volume in absolute terms, and not the mode
volume \emph{per molecule}.  This suggests  evanescently
confined radiation modes in plasmonic~\cite{Kim15} or phonon polariton~\cite{Caldwell13} systems may be a promising venue to explore this physics.

\subsection{Vibrational degrees of freedom}
\label{sec:vibr-degr-freed}

In the vibrational case, even without disorder, an exact solution of
Eq.~(\ref{eq:7}) via Fourier transformation is no longer possible,
because the vibrational degrees of freedom are modeled as quantum
degrees of freedom with their own quantum dynamics.  As such, they
break translational invariance --- i.e.\ localized vibrational
excitations can scatter between different polariton momentum states.
Thus, once again we must use the Schrieffer-Wolff formalism.  If we
start from Eq.~(\ref{eq:7}), solving the equation
$[\swmat,\hat{H}_0]=i\hat{H}_1$ is now more challenging than for the
rotational case, as $\hat{H}_0$ involves terms that couple the
electronic state to the vibrational quantum state.  The equation can however
be solved  in the form of a power series,
\begin{equation}
\swmat = \sum_{\vec{k},n,j}
\frac{-i\tilde{g}_{\vec{k},n} }{%
  (\tilde{\omega}_{\vec{k}} + \epsilon_n)^{j+1}}
 \left(
    \hat{O}_{j} 
\psic_{\vec{k}}\sigma^+_n e^{-i \vec k \cdot \vec r_n}    
    - \text{H.c.}\right),
\end{equation}
where the operators $\hat{O}_{j}$ are defined by the recursion relation
\begin{multline*}
  \hat{O}_{j} = \sum_m \Omega_m \biggl\{
  \left[\hat{O}_{j-1}, \bd_m \hb_m +
      \frac{\sqrt{S_m}}{2}(\bd_m + \hb_m) \right] 
    \\- \hat{O}_{j-1}
    \sqrt{S_m}(\hb_m+\bd_m) \biggr\},
\end{multline*}
with the base case $\hat{O}_0=1$.  This
expansion then allows one to write out the effective Hamiltonian in
the same form as above, but with the molecular Hamiltonian,
\begin{align}
  \hat{\tilde{h}}_{n} &= {\hat{h}}^{(0,\text{RWA})}_n
  -
  \left[\frac{1 - \sigma_z}{4} \right]
  \sum_{\vec{k},j} \frac{\tilde{g}_{\vec{k}}^2 
  }{(\tilde{\omega}_{\vec{k}} + \epsilon_n)^{j+1}} 
  (\hat{O}_{j} + \hat{O}_{j}^{\dagger}),
\end{align}
with ${\hat{h}}^{(0,\text{RWA})}_n$ the bare molecular Hamiltonian,
including the RWA coupling to light, and
vibrational terms of Eq.~(\ref{eq:7}).

This expression can be considered as a multinomial power series in the
quantities $\Omega_m/(\tilde{\omega}_{\vec{k}}+\epsilon_n)\ll 1$ for
each vibrational mode $m$.  For typical parameters, such quantities
are small, and so we may truncate at first order, i.e.\ keep terms up
to $j=1$. Beyond $j=1$, the expression becomes considerably more
complicated, as cross terms between different vibrational modes
appear.  Up to $j=1$, we find an effective molecular Hamiltonian which
we write out in full (neglecting constant terms):
\begin{multline}
  \label{eq:12}
  \hat{\tilde{h}}_{n} = \frac{\epsilon_n+K_0}{2}\sigma_z + 
  \sum_{\vec{k}} \tilde{g}_{\vec{k},n} 
  \left( {\hat \psi}^\dagger_{\vec{k}} e^{-i \vec{k} \cdot \vec{r}} \sigma^-_n
    + \text{H.c.} \right)
  \\
  +
  \sum_m \Omega_m
  \left(\bd_m b_m +\frac{\sqrt{S_m}}{2}(\bd_m + \hb_m)\sigma^z
  \right.\\ \left.
  +K_1 \left[\frac{{1 -  \sigma_z} }{2}\right]
     \sqrt{S_m}
    (b_m+\bd_m)  
  \right).
\end{multline}
The $j=0$ term gave an energy shift of two-level systems with the same
form $K_0$ found previously.  The $j=1$ term is on the last line, and
describes a shift to the vibrational modes. The coefficients for these
terms are defined by a generalization of that used in Eq.\
\eqref{eq:defK0} for the rotational case above,
\begin{displaymath}
  K_j(\epsilon) =
  \sum_{\vec{k}} 
  \frac{\tilde{g}_{\vec{k}}^2 }{(\tilde{\omega}_{\vec{k}} + \epsilon)^{j+1}},
\end{displaymath}
where we may find an analytic form for the resulting integral 
\begin{displaymath}
	K_j = \frac{f\epsilon\zeta}{2\pi l_c c^2 j}\left(\frac{1}{(\epsilon+\tilde\omega_0)^j} - \frac{1}{(\epsilon+\tilde\omega_\Lambda)^j}\right).
\end{displaymath}
We may once again note that none of the terms $K_j(\epsilon)$ are
proportional to the number of molecules $N_m$: vacuum-state molecular
adaptation is not collectively enhanced.

From the form of Eq.~(\ref{eq:12}) it is clear that the term in $K_1$
describes a reduction in the offset between the vibrational ground
state in the two electronic states, as the virtual excitations admix
the excited electronic state configuration into the ground state.
This can be viewed as a reduction of the Huang-Rhys parameter,
$S_{m,\text{eff}} = S_m [1 - 2 K_1]$. 
This point is illustrated in Fig.~\ref{fig:cartoon}.
 As noted earlier, the
dependence of $g_{\text{eff}}$ on the vibrational degrees of freedom
is more complicated: one must calculate the overlap between the
vibrational states of the ground and excited electronic state
manifolds.  The reason that the vibrational states differ in these
manifolds is the existence of the terms $(\bd_m + \hb_m)\sigma^z$,
which correspond to displacement of the vibrational coordinate
dependent on the electronic state of the molecule.  As such, a
reduction of $S_{m,\text{eff}}$ would lead to an enhanced matter-light
coupling.  However, using the same typical experimental values as
before we find that this dimensionless shift has a value of
approximately
\begin{math}
  K_1 = 2.9 \cdot 10^{-8}
\end{math}.  Thus, once again one may conclude that the characteristic
scale of any vibrational molecular adaptation (determined by $K_1$) is
negligibly small.

\subsection{Other aspects of molecular state}

The discussion so far has focused on two specific microscopic
mechanisms which might have led to self-consistent adaptation of the
molecules so as to enhance their coupling to light.  Here we note
those aspects of the above results which can be easily generalized to
other microscopic mechanisms.  Examples of such other potential
mechanisms include solvation of the molecule, molecular configuration, and
charge transfer state.  In some cases, these will require detailed
modeling of the specific process, particularly for degrees of freedom
with energy scales larger than temperature, where quantum effects
become important, as in the example of vibrational modes
above. However, the basic idea can be illustrated in the simplest
case, where some aspect of configuration can be parametrized by a
classical variable.

In the case where classical parametrization applies, the description
is very similar to the discussion of orientational degrees of freedom:
One considers a classical variable $x_n$ which parametrizes some
aspect the state for molecule $n$.  Associated with this will be an
energy function $\Lambda_n(x_n)$, and, for self-consistent adaptation
to occur, the matter-light coupling must depend on this variable as $g
\mapsto g \phi_n (x_n)$.  The same perturbative analysis as discussed
above will then lead to the effective energy function: $ E_n(x_n) =
\Lambda_n(x_n) - K_0 \phi_n(x_n)^2 $.  This in turn means that the
self-consistent effective matter-light coupling will take the form:
\begin{displaymath}
  \frac{g_{n,\text{eff}}^2}{g^2_n}
  = \frac{1}{\beta} \frac{d}{d K_0} \ln\left[
    \int dx_n e^{-\beta E_n(x_n)}
  \right].
\end{displaymath}
Since this involves the same sum $K_0$ as defined in
Eq.~(\ref{eq:defK0}), the same absence of scaling with number of
molecules occurs i.e.\ it is the single-molecular, rather than collective, coupling which is important.  Moreover, since $\beta K_0 \ll 1$, then:
\begin{displaymath}
  {g_{n,\text{eff}}^2}
  \simeq {g^2_n} \langle\langle \phi_n(x_n)^2 \rangle\rangle_0
  + \mathcal{O}(\beta K_0)
\end{displaymath}
where $\langle\langle \ldots \rangle\rangle_0$ indicates thermal
averaging with the bare energy function $\Lambda_n(x_n)$, i.e.\ any
such vacuum-state self-consistent adaptation of molecules is
suppressed by the small parameter $\beta K_0$.

\section{Vibrational dressing of the Exciton-Polariton spectrum}
\label{sec:vibrational-freedom}

In the previous section we have seen that while self-consistent
molecular adaptation due to matter-light coupling is possible, neither
the strength nor the dependence upon molecular concentration are
consistent with the results reported in
Ref.~\cite{Canaguier-Durand2013}.  In this section we show how directly calculating the absorption spectrum from the model of vibrationally dressed polaritons introduced in the previous section can lead to a rather different mechanism which could however be responsible for the
features observed.  This mechanism arises from the combination of
vibrational dressing of the spectrum and the effects of disorder. We focus in
particular on the peak in the spectrum near the bare excitonic
resonance.  We will see that the shape of this feature depends
strongly on the vibrational state of the molecules.

This section is divided into two subsections, in
section~\ref{sec:self-energy-incl} we first discuss how vibrational
excitations should be included in the calculation of the disordered
polariton spectrum.  This follows the method outlined in
appendix~\ref{sec:schr-wolff-transf}, and so all that is required is
to calculate the excitonic self energy.  We then discuss the resulting
form of the spectrum in section~\ref{sec:evol-absorpt-spectr}.

\subsection{Self energy of vibrationally dressed excitons}
\label{sec:self-energy-incl}

As discussed in appendix~\ref{sec:absorpt-transm-refl}, the
absorption, emission and transmission spectra can all be found in
terms of the photon retarded Green's function.  In this approach, all
the properties of the molecules (i.e.\ inhomogeneous broadening,
vibrational dressing etc.) are incorporated via the excitonic self
energy $\tilde{\Sigma}_{\vec{k},xx}(\nu)$.  We must therefore
calculate this quantity, defined in
Eq.~(\ref{eq:self-energy},\ref{eq:13}), for the vibrationally dressed
Hamiltonian, Eq.~(\ref{eq:7}).  To do this, it is useful to label the
eigenstates as $\vert p\,{\uparrow}\rangle, \vert
q\,{\downarrow}\rangle$, corresponding to the vibrational eigenstates
in the excited electronic state manifold and the ground electronic
state manifold.  We denote the energies of these states as $E_p^\uparrow$, and
$E_q^\downarrow$ and we introduce the overlap matrix element:
$\alpha_{pq} = \vert \langle p\,{\uparrow} \vert \sigma^+ \vert
{\downarrow}\, q \rangle \vert ^2$.  This matrix element describes the
extent to which the state with $q$ vibrational excitations in the
electronic ground manifold overlaps with the state in the excited
electronic manifold with $p$ excitations.  In order to find these
eigenvalues and eigenfunctions we must diagonalize the vibrational
problem, $\hat{H}_{\text{vib}}=\Omega\bd \hb + \Omega
\frac{\sqrt{S}}{2}\sigma^z \left(\bd +\hb \right)$ in the electronic
ground and excited states. As discussed in
Sec.~\ref{sec:vibr-degr-freed}, the renormalization of these
parameters due to virtual pair creation is very small, and so we
neglect it in the following.

In terms of these quantities we may write the self energy, including
inhomogeneous broadening of excitonic energies in the form:
\begin{align}
  \Sigma^{\text{RWA}}_{\vec{k},+-}(\nu)=- \frac{g_{\vec{k}}^2N_m}{\mathcal{Z}}   
\!\int_{-\infty}^{\infty} \!\!\!\! d\epsilon h(\epsilon)
\sum_{p,q}
\frac{\alpha_{pq}\left[e^{-\beta E_q^{\downarrow}}-e^{-\beta E_p^{\uparrow}}\right]}{%
  \nu+i\gamma +(E_q^{\downarrow}-E_p^{\uparrow})},
 \label{eq:RWA-self-energy}
\end{align}
where $h(\epsilon)$ is again the distribution of exciton energies,
 as in Sec.~\ref{sec:rotational-freedom}, and we have again
ignored the subleading effects of the exciton energy distribution on
the coupling strength, by using $g_\vec{k}^2 N_m \equiv \sum_n
g_{\vec{k},n}^2$.  It should be noted that the energies $E_p^\uparrow,
E_q^\downarrow$ depend (linearly) on the energy $\epsilon$, but that
the matrix elements $\alpha_{pq}$ are not dependent on this energy
scale.  

In the following, we model the inhomogeneous broadening of excitons by
using the distribution $h(\epsilon) \propto
\Theta(\epsilon)e^{(\epsilon - \epsilon_0)^2/2\sigma^2}$, i.e.\ a
truncated Gaussian distribution of excitonic energies.  The truncation
has little effect, but is formally required, as negative energy states
cannot physically exist, and cause problems in the ultrastrong
coupling limit~\cite{Ciuti2006, DeLiberato09}.

\subsection{Dependence of absorption spectrum on vibrational state}
\label{sec:evol-absorpt-spectr}

The exciton spectrum in the absence of vibrational dressing was shown
previously in Fig.~\ref{fig:abs_simple}.  Before exploring the effect
of vibrational modes, it is helpful to summarize how the residual
exciton peak arises.
Mathematically, the excitonic feature in the polaritonic spectrum can
be understood directly from the form of Eq.~(\ref{eq:3}) and the
definition of the absorption spectrum.  The peak at the exciton
frequency occurs because the imaginary part of the retarded Green's
function has a numerator involving the imaginary part of the self
energy, $\Im[\tilde{\Sigma}_{\vec{k},xx}(\nu)]$, and this self energy
has a peak at the excitonic energy.  However, the weight of the
residual excitonic peak reduces as the matter-light coupling increases,
because the excitonic self energy also appears, squared, in the
denominator.  It is also important to note that this residual
excitonic peak does not appear in the transmission spectrum: since
$T_{\vec{k}}(\nu) \propto \vert G^R_{\vec{k},xx}(\nu)\vert ^2$, there
is no term in the numerator of $T_{\vec{k}}(\nu)$ arising from the exciton self energy.
Thus, the residual exictonic peak is a feature of the absorption (and reflection), but
not the transmission spectrum.

\begin{figure}[htpb]
\hspace*{-0.2cm}
\includegraphics[width=3.2in]{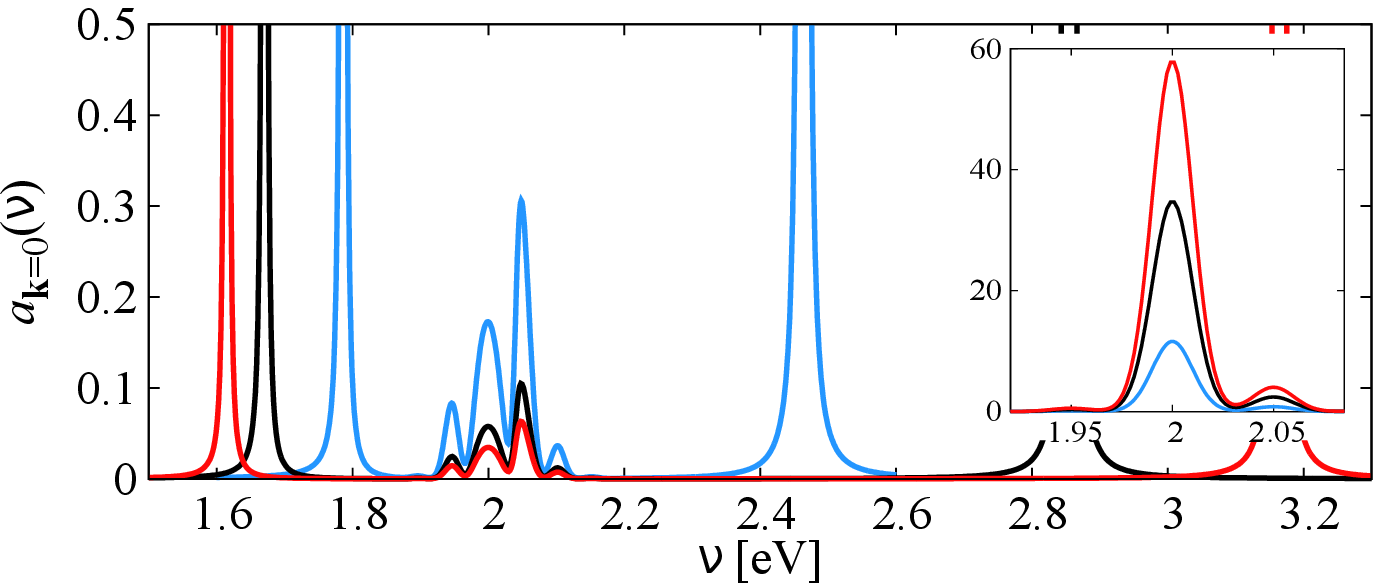}
\includegraphics[width=3.2in]{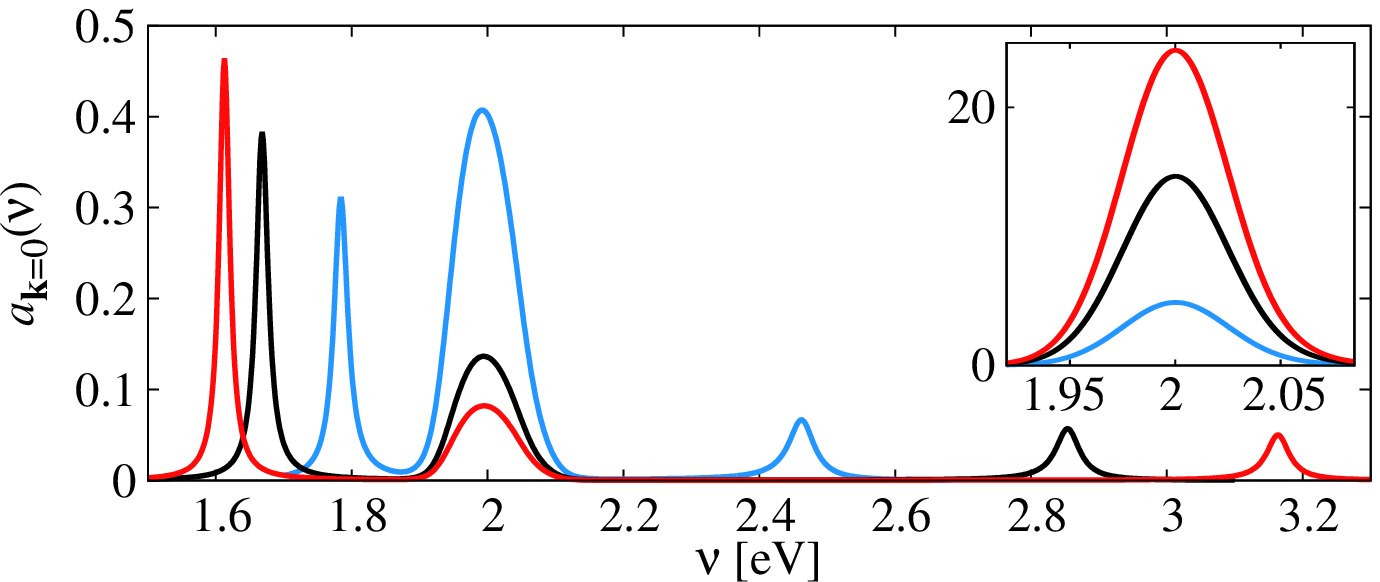}
\caption{(color online) The evolution of absorption spectra with the strength of
  $g\sqrt{N}$  for a system with a single
  vibrational mode with $S=0.06, \Omega=0.05$eV. The insets show
  the equivalent absorption spectra for bare molecules. The other parameters are as in Fig.~\ref{fig:abs_simple}. Panel (a) shows the spectrum for a perfect cavity while panel (b) includes the effects of cavity losses and non-radiative excitonic decay exactly as in Fig.~\ref{fig:abs_simple}(b).}
\label{fig:abs_vs_g}
\end{figure}

Figure~\ref{fig:abs_vs_g} shows an equivalent set of spectra to
Fig.~\ref{fig:abs_simple}, but with vibrational dressing.  For
comparison in the insets we show the absorption spectra of a bare
molecule $\Im[\tilde{\Sigma}_{\vec{k},xx}(\nu)]$, i.e.\ the spectra
without a cavity. Panel (a) shows the simple case where cavity losses are ignored, while panel (b) shows the more experimentally relevant case where these effects are included (discussed in appendix~\ref{sec:absorpt-transm-refl}). In the following we use the notation $p-q$ which denotes transitions in the molecule from the state with $p$ vibrational excitations in the electronic ground state to the state with $q$ vibrational excitations in the electronic excited state. The presence of the vibrational modes causes
dramatic changes to the residual excitonic peak in the polariton
spectrum not seen in the bare excitonic absorption. For the bare
molecule the spectral weight associated with transition with the
``zero phonon line'', i.e.\ the transition denoted
$0-0$ in the notation introduced above is completely dominant, only a small amount of weight visible in the sideband which corresponds to $0-1$ transition.  When the molecule is placed
inside a cavity, the vibrational sidebands corresponding to the $1-0$
and $0-1$ transitions become much more prominent.  Physically this is
because the spectral weight that had been associated with the $0-0$
transition in the bare molecular spectrum 
 has been moved into the polariton peaks of the spectrum.  
The mathematical form of the Green's function makes clear that
formation of the polariton spectral feature is predominantly at the expense
of whatever feature dominates the excitonic emission spectrum; here this
is the $0-0$ feature.
The excitonic
feature corresponds to the ``left-over'' spectral weight associated with the subradiant states.  Thus, the
vibrational sidebands have been ``excavated'' by removing the dominant
feature from the zero-phonon line. Including the the effects of cavity losses and non-radiative excitonic decays, as can be seen in Fig.~\ref{fig:abs_vs_g}~(b), washes out this complex sideband structure of the central peaks and the spectrum as a function of coupling strength looks very similar to that obtained  without coupling to vibrational modes, as in Fig.~\ref{fig:abs_simple}. However as we discuss below, there are still effects which can be observed which are a direct consequence of the vibrational structure. 

\begin{figure}[htpb]
\hspace*{-0.2cm}
\includegraphics[width=3.2in]{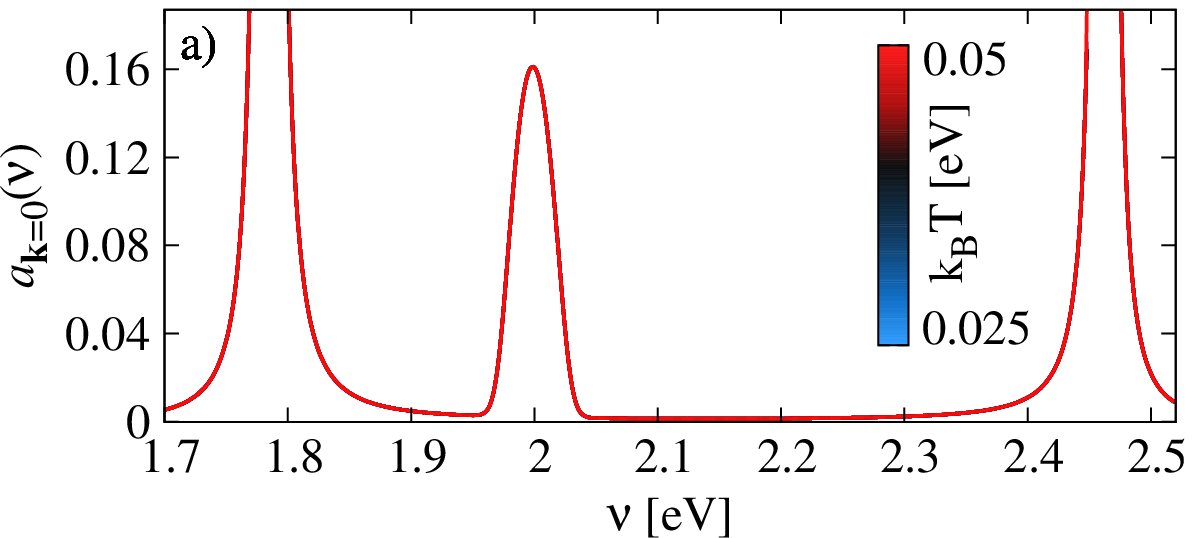} 
\includegraphics[width=3.2in]{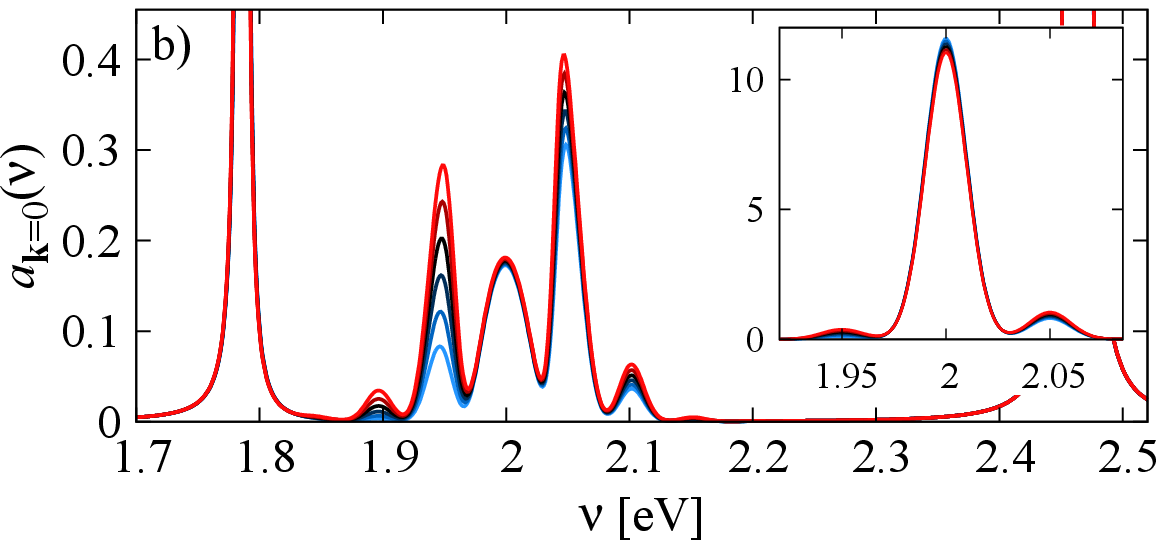}
\includegraphics[width=3.2in]{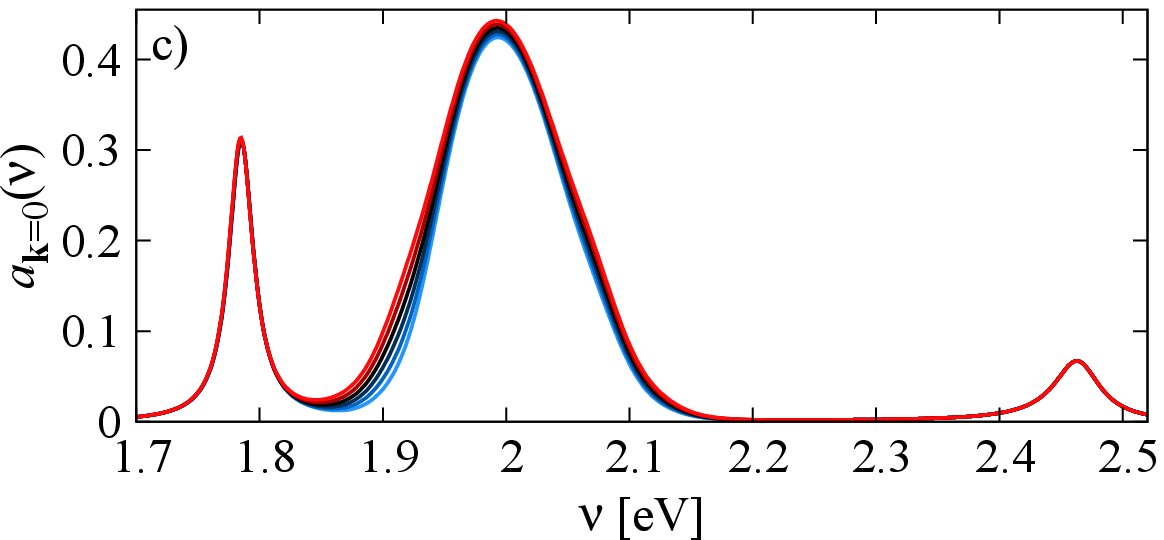}
\caption{(color online) Evolution of absorption spectra with the temperature (values
  shown on the colorscale) for: (a) A system with no coupling to
  vibrational modes, $S=0$, (b) Coupling to
  a single vibrational mode with $S=0.02, \Omega=0.05$eV. The
  inset shows the bare molecular spectrum. (c) Including the effects of cavity losses and non-radiative excitonic decay. All panels are for
  $g\sqrt{N}=0.3$eV, and other parameters as in previous figures.}
\label{fig:abs_vs_T}
\end{figure}

Fig.~\ref{fig:abs_vs_T} illustrates the evolution of spectra with
temperature. Panel (a) shows that without coupling to vibrational
modes there is no notable temperature dependence. In the presence of
the vibrational dressing, a strong temperature dependence appears.  At
higher temperatures, there is a greater thermal occupation of the
vibrational modes hence the spectral weight under the vibrational
peaks rises.  While this has a small effect in the bare molecular
spectrum (where the $0-0$ transition dwarfs all other features), see inset in (b), it is
very pronounced in the polariton spectrum and is even visible in the presence of large cavity losses as in Fig.~\ref{fig:abs_vs_T}(c).

\begin{figure}[thpb]
\includegraphics[width=3.2in]{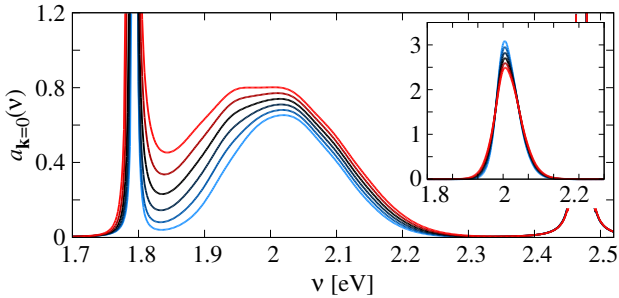}
\caption{(color online) Temperature dependence of absorption spectrum for a system
  with strong disorder $\sigma=0.025eV$. The main panel shows the
  optical spectrum of the strongly coupled system, the inset shows the
  spectrum of the bare excitons.  In this case, we include two
  vibrational modes: $S_1=0.2, \Omega_1=0.05$eV, and $S_2=0.3,
  \Omega_2=0.0$4eV.  Other parameters as in Fig.~\ref{fig:abs_vs_T}.}
\label{fig:abs_large_disorder}
\end{figure}

The figures so far have shown results where disorder is relatively
small, and so vibronic replicas can be clearly observed for the good
cavity, but merge for the bad cavity limit.  To show that
small disorder is
 not required for the strong temperature dependence to occur
Fig.~\ref{fig:abs_large_disorder} shows the effects of
large disorder (i.e. inhomogeneous broadening).  Just as seen for
the homogeneous broadening in Fig.~\ref{fig:abs_vs_T}(c), a
temperature dependence of the residual excitonic peak
is still visible.  In this figure, we have
also included a more complicated vibrational spectrum, involving two
vibrational modes. Such a system will exhibit behavior similar to that
of a single mode but with a large vibrational coupling~\footnote{One
  may note that if the vibrational modes were degenerate
  $\Omega_1=\Omega_2=\Omega$, then the effective Huang-Rhys parameter
  will be $S_{\text{eff}}=S_1+S_2$}.

\section{Discussion}
\label{sec:discussion}

In this paper we have presented two microscopic models which could in
principle describe self-consistent molecular adaptation so as to maximize
the vacuum-state coupling to light.  In both cases, the crucial feature of
the model is the counter-rotating terms in the matter-light coupling.
These allow virtual fluctuations in the ground state, that lower the
ground state energy depending on the configuration of the molecules.
This energy gain is the only energy gain that can be relevant in the
linear response regime --- i.e.\ the question of whether the excited
states would have lower energy is not of relevance while the
system is only weakly pumped.  We found that while such a mechanism
for molecular adaptation does exist, it does not show any collective
enhancement, in contrast to the polariton splitting, and does not
therefore lead to significant molecular adaptation, even when the polariton
splitting $g_{\vec{k}}\sqrt{N_m} \sim \omega, \epsilon$.

The appearance of a residual exciton peak in the polariton spectrum
would be affected by any such  self-consistent molecular adaptation if its scale
were sufficient.  However, for relevant parameters, such effects are
dwarfed by a far more dramatic effect, of vibrational dressing of the
residual exciton peak.  This leads to a pronounced temperature
dependence of the feature near the exciton energy in the absorption
spectrum.

While the  molecular adaptation energy scale is not collectively
enhanced, the basic underlying physics could potentially be relevant
in single molecule strong coupling, e.g.\ with plasmonic
resonances~\cite{torma15}.  In such cases, rather than having many
molecules in a large mode volume, the mode volume is reduced so that
strong, or even ultra-strong coupling occurs at the single molecule
level, so that both the polariton energy and the 
molecular adaptation energy become large. In such a case, one may hope
to see either reorientation, or renormalization of the Huang-Rhys
parameter due to strong coupling.

Another intriguing direction for future research is to consider how
the physics discussed in this manuscript interacts with the physics of
polariton condensation and lasing~\cite{Carusotto2013}.  Polariton
condensation has been seen in both
inorganic~\cite{Kasprzak2006,snoke07science} and
organic~\cite{Kena-Cohen2008,Forrest2010,Plumhof2013,Daskalakis2014}
systems.  In addition, condensation of photons has been seen for
weakly coupled systems of organic molecules~\cite{Klaers2010}.
Theoretical
work~\cite{Litinskaya2004,Litinskaya2006,Litinskaya2008,Michetti2009,Fontanesi2009,Mazza2009,Mazza2013,Cwik2014,Michetti2015}
has begun to address some of the peculiarities of the organic polariton
system, including effects of disorder and of vibrational modes.
However, features as seen in this paper, resulting from the interplay
of these may lead to further exotic behavior in the high density
condensed phase.

\begin{acknowledgments}
  We are grateful for comments from T.~Ebbesen on an earlier version
  of this paper.  JK acknowledges helpful discussions with David
  Lidzey and Brendon Lovett.  JAC acknowledges support from EPSRC. JK
  and PGK acknowledge financial support from EPSRC program ``TOPNES''
  (EP/I031014/1).  JK acknowledges support from the Leverhulme Trust
  (IAF-2014-025).  SDL is Royal Society Research Fellow. SDL
  acknowledges financial support from EPSRC grant EP/M003183/1. PGK
  acknowledges support from EPSRC grant EP/M010910/1.  JK, PGK and SDL
  acknowledge support from the British Council for the meeting which
  initiated this work.
\end{acknowledgments}

\paragraph*{Note added:} During the final preparation of this manuscript,
another paper~\cite{Galego2015} appeared, also reporting the fact that
ground-state bond length depends on the single-molecule coupling $g$, not
the collective coupling $g\sqrt{N_m}$.

\appendix

\section{Absorption, Transmission and Reflection spectrum of exciton-polariton system}
\label{sec:absorpt-transm-refl}

In this appendix we summarize the calculation of the absorption,
transmission and reflection spectra.  Some subtleties arise because we
wish to calculate the spectrum of a model with ultrastrong coupling,
i.e.\ without making the rotating wave approximation.  Such results
were first calculated by~\citet{Ciuti2006}, here we present a synopsis
of these results, as well as a ``dictionary'' to translate the results
of that paper into the language of Green's functions.  We begin by
defining the retarded Green's function~\cite{agd}.  Because we
consider both co- and counter-rotating terms, we must consider both
normal and anomalous Green's functions. i.e.\ we must include number
non-conserving terms which appear for ultrastrong coupling, and thus
we consider a matrix Green's function:
\begin{displaymath}
  G^R_{\vec{k}}(t,t^\prime) = -i 
  \begin{pmatrix}
    \langle [ \psi^{}_{\vec{k}}(t), \psi^\dagger_{\vec{k}}(t^\prime) ] \rangle
    &
    \langle [ \psi^{\dagger}_{-\vec{k}}(t), \psi^\dagger_{\vec{k}}(t^\prime) ] \rangle
    \\
    \langle [ \psi^{}_{\vec{k}}(t), \psi^{}_{-\vec{k}}(t^\prime) ] \rangle
    &
    \langle [ \psi^{\dagger}_{-\vec{k}}(t), \psi^{}_{-\vec{k}}(t^\prime) ] \rangle
  \end{pmatrix}
\end{displaymath}
\begin{widetext}
  In terms of the Bogoliubov transformed operators, i.e.\ the operators
  appearing in Eq.~(\ref{eq:2}) the inverse Green's function takes the
  form
\begin{displaymath}
  [\tilde{G}^R_{\vec{k}}(\nu)]^{-1} = 
  \begin{pmatrix}
    \nu + i \tilde{\kappa}(\nu) - \tilde{\omega}_{\vec{k}} + \tilde{\Sigma}^{}_{\vec{k},xx}(\nu) 
    &
    + i \tilde{\kappa}(\nu) + \tilde{\Sigma}_{\vec{k},xx}(\nu) 
    \\
    - i \tilde{\kappa}^\ast(-\nu) + \tilde{\Sigma}^\ast_{\vec{k},xx}(-\nu) 
    &
    -\nu - i \tilde{\kappa}^\ast(-\nu) - \tilde{\omega}_{\vec{k}} + \tilde{\Sigma}^\ast_{\vec{k},xx}(-\nu) 
  \end{pmatrix},
\end{displaymath}
where $\tilde{\Sigma}_{\vec{k},xx}^{}(\nu)$ is the self energy for a
photon of in-plane momentum $\vec{k}$, arising from the excitonic
response (discussed further below), and $\tilde{\kappa}(\nu)$ is the
loss rate.
\end{widetext}
Here, following~\cite{Ciuti2006} we have used a frequency dependent
complex loss rate $\tilde{\kappa}(\nu)$.  Frequency dependence is
required for physical consistency in the case of ultrastrong coupling
--- Markovian loss and ultrastrong coupling would predict a perpetual
light source.  Frequency dependent loss requires, via the
Kramers-Kronig relation, a corresponding Lamb shift, which is
incorporated into the imaginary part of $\tilde{\kappa}(\nu)$.  Both
$\tilde{\Sigma}$ and $\tilde{\kappa}$ are written for the Bogoliubov
transformed operators, and so both these terms incorporate a pre-factor
$\omega_{\vec{k}}/\tilde{\omega}_{\vec{k}}$ to account for the
Bogoliubov transformation of the combination $(\hat \psi^{}_{\vec{k}}
+ \hat \psi^\dagger_{-\vec{k}})$.

The specific self energy required,
$\tilde{\Sigma}_{\vec{k},xx}^{}(\nu)$, corresponds to correlation
functions of the $\hat{\sigma}^x$ excitonic operators.  In the absence
of strong (i.e.\ beyond RWA) excitonic damping (see~\cite{Ciuti2006} for the more general case),
this self energy can however be related to results in the rotating wave approximation by:
\begin{equation}
  \tilde{\Sigma}_{\vec{k},xx}(\nu) =
  \frac{\omega_k}{\tilde{\omega}_k}\left(\Sigma^{\text{RWA}}_{\vec{k},+-}(\nu)
    + \left[\Sigma^{\text{RWA}}_{\vec{k},+-}(-\nu) \right]^{\star} \right),
  \label{eq:self-energy}
\end{equation}
where the
 expression $\Sigma^{\text{RWA}}_{\vec{k},+-}$ is the ``standard'' self energy
that would appear in the rotating wave approximation, depending on the
correlation of ${\hat \sigma}^+, {\hat \sigma}^-$ operators.  These
can most easily be found by analytic continuation from imaginary time
to real time, starting from the Matsubara self energy,
\begin{multline}
  \label{eq:13}
  \Sigma^{\text{RWA}}_{\vec{k},+-}(i\omega_m) 
  = 
  \frac{1}{\mathcal{Z}} 
  \sum_n g_{\vec{k},n}^2
  \int_{0}^{\beta} \!\!d\tau e^{-i\omega_m \tau}
  \\\times
  \sum_{p}
  \langle p\vert 
  \hat{\sigma}_n^+(\tau)\hat{\sigma}_n^-(0)
  \vert p \rangle e^{-\beta E_p} 
\end{multline}
and replacing the Matsubara frequency by $i\omega_m \to \nu + i 0^+$.
The sum over $p$ appearing here is over all states of the excitonic
system, and $\mathcal{Z}$ is the partition function.  For the
``vacuum'' state we consider --- i.e.\ in the absence of strong
pumping --- these are the bare exciton states, including the
quantum states of any auxiliary degrees of freedom.

Using the input-output formalism~\cite{Collett1984} adapted to the
ultrastrong coupling regime~\cite{Ciuti2006}, one can write a
frequency dependent scattering matrix relating input and output
fields at the left and right sides of the cavity,
\begin{displaymath}
  \mathcal{S}_{\vec{k}}(\nu)
  =
  \begin{pmatrix}
    1 - i \phi_{\vec{k},L}(\nu) &
    - i \sqrt{\phi_{\vec{k},L}(\nu) \phi_{\vec{k},R}(\nu)}
    \\
    - i \sqrt{\phi_{\vec{k},L}(\nu) \phi_{\vec{k},R}(\nu)}
    &
    1 - i \phi_{\vec{k},R}(\nu)
  \end{pmatrix}
\end{displaymath}
where $\phi_{\vec{k},\sigma}(\nu)= \kappa_\sigma^\prime(\nu)
G^R_{\vec{k},xx}(\nu)$ with $\kappa_{L,R}^\prime(\nu)$ are the real
parts of the loss rates arising from the left and right mirrors and
the quantity $G^R_{\vec{k},xx}(\nu)$ relates to the matrix retarded Green's
function as $G^R_{\vec{k},xx}(\nu) = I^T G^R_{\vec{k}}(\nu) I$ where
$I^T= \begin{pmatrix} 1 & 1 \end{pmatrix}$.  This structure means the
Bogoliubov transformation corresponds to $G^R_{\vec{k},xx}(\nu)
=(\omega_{\vec{k}}/\tilde{\omega_{\vec{k}}}) \tilde{G}^R_{\vec{k},xx}(\nu)$
and the Bogoliubov transformed Green's function takes the form:
\begin{equation}
  \label{eq:11}
  \tilde{G}^R_{\vec{k},xx}(\nu)  =
  \frac{2 \tilde{\omega}_{\vec{k}}}{%
    \nu^2 - \tilde{\omega}_{\vec{k}}^2 
    + 2\tilde{\omega}_{\vec{k}} \tilde{\Sigma}_{\vec{k},xx}(\nu)
    + i\tilde{\omega}_{\vec{k}} \tilde{\kappa}(\nu)}.
\end{equation}
One can then find the transmission $T_{\vec{k}}(\nu)$, reflection $R_{\vec{k}}(\nu)$
and absorption $A_{\vec{k}}(\nu)$ coefficients by considering the modulus
square of various coefficients.  Clearly $T_{\vec{k}}(\nu)=
\kappa_L^\prime(\nu)\kappa_R^\prime(\nu) \vert G^R_{\vec{k},xx}(\nu)\vert ^2$ is
independent of which direction light is incident from, while the
absorption coefficient $A_{\vec{k},\sigma=L,R}(\nu)$ takes the form
\begin{equation}
  \label{eq:15}
  A_{\vec{k},\sigma}(\nu) = -  \kappa^\prime_\sigma(\nu) \left[
    2 \Im\left[G^R_{\vec{k},xx}(\nu)\right] 
    + \kappa(\nu) \left\vert G^R_{\vec{k},xx}(\nu)\right\vert ^2 \right],
\end{equation}
with $\kappa(\nu)=(\kappa_L(\nu)+\kappa_R(\nu))/2$. The prefactor in
this expression shows the obvious dependence on the transmissivity of
the input mirrors.

In order to separate mirror transmissivity dependent features from the
``intrinsic'' properties of the ultrastrong coupling we will consider
below the two quantities $t_{\vec{k}}(\nu)= \vert G^R_{\vec{k},xx}(\nu)\vert ^2$
as being proportional to the transmission, and $a_{\vec{k}}(\nu) = -2
\Im[G^R_{\vec{k},xx}(\nu)]$ as controlling the absorption in the limit
of a good cavity, i.e.\ $\kappa(\nu) \to 0$.  The quantity
$a_{\vec{k}}(\nu)$ differs from the full absorption as it neglects
interference effects at the input mirror.  

\begin{figure}[htpb]
  \centering
\includegraphics[width=3.2in]{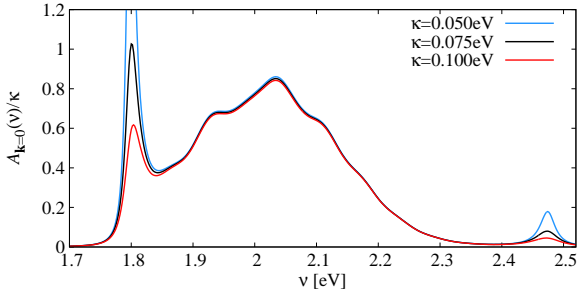}
  \caption{(color online) Absorption spectrum, as defined by the full expression,
    Eq.~(\ref{eq:15}), plotted for $S_1=0.25$, $S_2=0.35$, 
    $\Omega_1=0.06$eV, $\Omega_2=0.05$eV, $k_BT=0.035$eV and various cavity loss rates, $\kappa$. 
    Other parameters are as in Fig.~\ref{fig:abs_large_disorder}. 
    As discussed in the text, in order that the polaritonic
    peaks have a finite width, one must include the effects of
    excitonic absorption or non-mirror cavity losses.  For this figure,
    we include an excitonic linewidth $\gamma=10^{-4}$eV.}
  \label{fig:true-absorption}
\end{figure}

For comparison to the results in Ref.~\cite{Canaguier-Durand2013}
where silver mirrors were used, we present in
Figures~\ref{fig:abs_simple}, \ref{fig:abs_vs_g}, and \ref{fig:abs_vs_T}
the effects of large cavity linewidth.  
Figure~\ref{fig:true-absorption} also shows how
the full absorption spectrum given by
Eq.~(\ref{eq:15}) evolves with varying linewidth.  
In calculating these spectra for large linewidth, an 
issue arises regarding the form of the
absorption spectrum: The equations written above assume that the only
photon loss is due to escape through the mirrors.  This means that the
$\kappa(\nu)$ appearing explicitly in Eq.~(\ref{eq:15}) (describing
interference effects from the mirror) is the same as the $\kappa(\nu)$
appearing in the denominator of the photon Green's function,
Eq.~(\ref{eq:11}), and one may check that for any frequency where
$\Im[\Sigma_{\vec{k},xx}(\nu)]$ is small, this causes a near
cancellation between the two contributions.  For the Gaussian exciton
density of states $h(\epsilon)$ used in 
this paper, this
cancellation almost completely suppresses the polariton peaks.
Such a cancellation in the absorption spectrum can be expected on
physical grounds: if the only loss channel for photons is the
mirrors, then there is no absorption. All photons that enter
eventually leave.  In a real device there are other photon loss
sources (absorbers, scattering by surface roughness).  Similarly, in a
real device, the excitons have a non-zero rate of non-radiative decay.
The effect of this is included by retaining a non-zero value of $\gamma$ in the denominator of the self-energy, Eq.~\eqref{eq:RWA-self-energy}. 
This leads to Lorentzian tails of $\Im[\Sigma_{\vec{k},xx}(\nu)]$,
giving a finite weight to the polariton peak in the absorption
spectrum; such an effect was included by~\citet{houdre96}.  We follow
this approach in plotting Fig.~\ref{fig:true-absorption}.

\section{Schrieffer-Wolff transformation}
\label{sec:schr-wolff-transf}

In section~\ref{sec:rotational-freedom} we make use of the
Schrieffer-Wolff approximation; for completeness we provide here a
brief explanation of this formalism.  The approach is based on dividing the Hamiltonian into two parts,
$\hat H = \hat H_0 + \hat H_1$, where the term $\hat H_1$ takes one
between different ``sectors''.  In our case, these sectors correspond
to different numbers of polaritons --- i.e.\ $H_1$ is the
``counter-rotating'' part of the Hamiltonian which simultaneous
creates a photon and excites a molecule.  

The aim of the Schrieffer-Wolff formulation is to make a unitary
transformation $\hat{\tilde{H}} = e^{i\swmat} \hat{H} e^{-i\swmat}$
such that the transformed Hamiltonian no longer has any coupling
between sectors.  Physically, this corresponds to eliminating the
effect of virtual pair creation and destruction, and deriving how such
virtual processes renormalize the Hamiltonian within a given sector.

If the Hamiltonian $\hat{H}_1$ can be treated perturbatively by
replacing $\hat{H}_1 \to \eta \hat{H}_1$ with $\eta$ a small
parameter, then one can consider a series solution $\swmat =
\sum_{n=1}^\infty \eta^n \swmat_{(n)}$.  In order to make the
first-order terms in $\eta$ vanish, one must choose
$[\swmat_{(1)},\hat{H}_0]=i\hat{H}_1$.
This then leads (setting $\eta=1$) to the expression:
\begin{align}
  \tilde{\hat{H}}
  = \hat{H}_0 + 
\frac{i}{2}\left[\swmat_{(1)},\hat{H}_1 \right] + 
\text{H.o.t}
\end{align}
where the higher order terms involve all $\swmat_{(n>1)}$.  Stopping
at leading order gives the expression in Eq.~(\ref{eq:8}),
corresponding to the leading order effects of virtual pair creation
and annihilation.  To solve $[\swmat_,\hat{H}_0]=i\hat{H}_1$ in
practice is straightforward if one knows the eigenspectrum of $H_0 =
\sum_n E^{(0)}_n \vert n\rangle \langle n|$,  which then allows one to write
\begin{displaymath}
  \swmat = i \sum_{n,m} 
  \frac{\vert n\rangle\langle n \vert \hat{H}_1 \vert m \rangle%
    \langle m\vert }{E^{(0)}_m-E^{(0)}_n}.
\end{displaymath}


\begin{thebibliography}{58}%
\makeatletter
\providecommand \@ifxundefined [1]{%
 \@ifx{#1\undefined}
}%
\providecommand \@ifnum [1]{%
 \ifnum #1\expandafter \@firstoftwo
 \else \expandafter \@secondoftwo
 \fi
}%
\providecommand \@ifx [1]{%
 \ifx #1\expandafter \@firstoftwo
 \else \expandafter \@secondoftwo
 \fi
}%
\providecommand \natexlab [1]{#1}%
\providecommand \enquote  [1]{``#1''}%
\providecommand \bibnamefont  [1]{#1}%
\providecommand \bibfnamefont [1]{#1}%
\providecommand \citenamefont [1]{#1}%
\providecommand \href@noop [0]{\@secondoftwo}%
\providecommand \href [0]{\begingroup \@sanitize@url \@href}%
\providecommand \@href[1]{\@@startlink{#1}\@@href}%
\providecommand \@@href[1]{\endgroup#1\@@endlink}%
\providecommand \@sanitize@url [0]{\catcode `\\12\catcode `\$12\catcode
  `\&12\catcode `\#12\catcode `\^12\catcode `\_12\catcode `\%12\relax}%
\providecommand \@@startlink[1]{}%
\providecommand \@@endlink[0]{}%
\providecommand \url  [0]{\begingroup\@sanitize@url \@url }%
\providecommand \@url [1]{\endgroup\@href {#1}{\urlprefix }}%
\providecommand \urlprefix  [0]{URL }%
\providecommand \Eprint [0]{\href }%
\providecommand \doibase [0]{http://dx.doi.org/}%
\providecommand \selectlanguage [0]{\@gobble}%
\providecommand \bibinfo  [0]{\@secondoftwo}%
\providecommand \bibfield  [0]{\@secondoftwo}%
\providecommand \translation [1]{[#1]}%
\providecommand \BibitemOpen [0]{}%
\providecommand \bibitemStop [0]{}%
\providecommand \bibitemNoStop [0]{.\EOS\space}%
\providecommand \EOS [0]{\spacefactor3000\relax}%
\providecommand \BibitemShut  [1]{\csname bibitem#1\endcsname}%
\let\auto@bib@innerbib\@empty
\bibitem [{\citenamefont {Hopfield}(1958)}]{hopfield58}%
  \BibitemOpen
  \bibfield  {author} {\bibinfo {author} {\bibfnamefont {J.}~\bibnamefont
  {Hopfield}},\ }\href {http://prola.aps.org/abstract/PR/v112/i5/p1555\_1}
  {\bibfield  {journal} {\bibinfo  {journal} {Phys. Rev.}\ }\textbf {\bibinfo
  {volume} {112}},\ \bibinfo {pages} {1555} (\bibinfo {year}
  {1958})}\BibitemShut {NoStop}%
\bibitem [{\citenamefont {Pekar}(1958)}]{pekar58}%
  \BibitemOpen
  \bibfield  {author} {\bibinfo {author} {\bibfnamefont {S.}~\bibnamefont
  {Pekar}},\ }\href {http://adsabs.harvard.edu/abs/1958JETP....6..785P}
  {\bibfield  {journal} {\bibinfo  {journal} {Sov. J. Exp. Theor. Phys.}\
  }\textbf {\bibinfo {volume} {6}},\ \bibinfo {pages} {785} (\bibinfo {year}
  {1958})}\BibitemShut {NoStop}%
\bibitem [{\citenamefont {Purcell}(1946)}]{purcell46}%
  \BibitemOpen
  \bibfield  {author} {\bibinfo {author} {\bibfnamefont {E.~M.}\ \bibnamefont
  {Purcell}},\ }\href@noop {} {\bibfield  {journal} {\bibinfo  {journal} {Phys.
  Rev.}\ }\textbf {\bibinfo {volume} {69}},\ \bibinfo {pages} {681} (\bibinfo
  {year} {1946})}\BibitemShut {NoStop}%
\bibitem [{\citenamefont {Bj\"{o}rk}\ \emph {et~al.}(1991)\citenamefont
  {Bj\"{o}rk}, \citenamefont {Machida}, \citenamefont {Yamamoto},\ and\
  \citenamefont {Igeta}}]{Bjork1991}%
  \BibitemOpen
  \bibfield  {author} {\bibinfo {author} {\bibfnamefont {G.}~\bibnamefont
  {Bj\"{o}rk}}, \bibinfo {author} {\bibfnamefont {S.}~\bibnamefont {Machida}},
  \bibinfo {author} {\bibfnamefont {Y.}~\bibnamefont {Yamamoto}}, \ and\
  \bibinfo {author} {\bibfnamefont {K.}~\bibnamefont {Igeta}},\ }\href
  {\doibase 10.1103/PhysRevA.44.669} {\bibfield  {journal} {\bibinfo  {journal}
  {Phys. Rev. A}\ }\textbf {\bibinfo {volume} {44}},\ \bibinfo {pages} {669}
  (\bibinfo {year} {1991})}\BibitemShut {NoStop}%
\bibitem [{\citenamefont {Weisbuch}\ \emph {et~al.}(1992)\citenamefont
  {Weisbuch}, \citenamefont {Nishioka}, \citenamefont {Ishikawa},\ and\
  \citenamefont {Arakawa}}]{weisbuch92}%
  \BibitemOpen
  \bibfield  {author} {\bibinfo {author} {\bibfnamefont {C.}~\bibnamefont
  {Weisbuch}}, \bibinfo {author} {\bibfnamefont {M.}~\bibnamefont {Nishioka}},
  \bibinfo {author} {\bibfnamefont {A.}~\bibnamefont {Ishikawa}}, \ and\
  \bibinfo {author} {\bibfnamefont {Y.}~\bibnamefont {Arakawa}},\ }\href
  {\doibase 10.1103/PhysRevLett.69.3314} {\bibfield  {journal} {\bibinfo
  {journal} {Phys. Rev. Lett.}\ }\textbf {\bibinfo {volume} {69}},\ \bibinfo
  {pages} {3314} (\bibinfo {year} {1992})}\BibitemShut {NoStop}%
\bibitem [{\citenamefont {Agranovich}(1968)}]{AgranovichRussian}%
  \BibitemOpen
  \bibfield  {author} {\bibinfo {author} {\bibfnamefont {V.~M.}\ \bibnamefont
  {Agranovich}},\ }\href@noop {} {\emph {\bibinfo {title} {{The Theory of
  Excitons}}}}\ (\bibinfo  {publisher} {Nauka},\ \bibinfo {address} {Moscow},\
  \bibinfo {year} {1968})\BibitemShut {NoStop}%
\bibitem [{\citenamefont {Davydov}(1971)}]{Davydov1971}%
  \BibitemOpen
  \bibfield  {author} {\bibinfo {author} {\bibfnamefont {A.~S.}\ \bibnamefont
  {Davydov}},\ }\href@noop {} {\emph {\bibinfo {title} {{Theory of Molecular
  Excitons}}}}\ (\bibinfo  {publisher} {Plenum Press},\ \bibinfo {address} {New
  York},\ \bibinfo {year} {1971})\BibitemShut {NoStop}%
\bibitem [{\citenamefont {Agranovich}(2009)}]{Agranovich2009a}%
  \BibitemOpen
  \bibfield  {author} {\bibinfo {author} {\bibfnamefont {V.~M.}\ \bibnamefont
  {Agranovich}},\ }\href@noop {} {\emph {\bibinfo {title} {{Excitations in
  Organic Solids}}}}\ (\bibinfo  {publisher} {Oxford University Press},\
  \bibinfo {address} {Oxford},\ \bibinfo {year} {2009})\BibitemShut {NoStop}%
\bibitem [{\citenamefont {Lidzey}\ \emph {et~al.}(1998)\citenamefont {Lidzey},
  \citenamefont {Bradley}, \citenamefont {Skolnick}, \citenamefont {Virgli},
  \citenamefont {Walker}, \citenamefont {Whittaker},\ and\ \citenamefont
  {Virgili}}]{lidzey98}%
  \BibitemOpen
  \bibfield  {author} {\bibinfo {author} {\bibfnamefont {D.~G.}\ \bibnamefont
  {Lidzey}}, \bibinfo {author} {\bibfnamefont {D.~D.~C.}\ \bibnamefont
  {Bradley}}, \bibinfo {author} {\bibfnamefont {M.~S.}\ \bibnamefont
  {Skolnick}}, \bibinfo {author} {\bibfnamefont {T.}~\bibnamefont {Virgli}},
  \bibinfo {author} {\bibfnamefont {S.}~\bibnamefont {Walker}}, \bibinfo
  {author} {\bibfnamefont {D.~M.}\ \bibnamefont {Whittaker}}, \ and\ \bibinfo
  {author} {\bibfnamefont {T.}~\bibnamefont {Virgili}},\ }\href {\doibase
  10.1038/25692} {\bibfield  {journal} {\bibinfo  {journal} {Nature}\ }\textbf
  {\bibinfo {volume} {395}},\ \bibinfo {pages} {53} (\bibinfo {year}
  {1998})}\BibitemShut {NoStop}%
\bibitem [{\citenamefont {Lidzey}\ \emph {et~al.}(1999)\citenamefont {Lidzey},
  \citenamefont {Bradley}, \citenamefont {Virgili}, \citenamefont {Armitage},
  \citenamefont {Skolnick},\ and\ \citenamefont {Walker}}]{lidzey99}%
  \BibitemOpen
  \bibfield  {author} {\bibinfo {author} {\bibfnamefont {D.~G.}\ \bibnamefont
  {Lidzey}}, \bibinfo {author} {\bibfnamefont {D.~D.~C.}\ \bibnamefont
  {Bradley}}, \bibinfo {author} {\bibfnamefont {T.}~\bibnamefont {Virgili}},
  \bibinfo {author} {\bibfnamefont {A.}~\bibnamefont {Armitage}}, \bibinfo
  {author} {\bibfnamefont {M.~S.}\ \bibnamefont {Skolnick}}, \ and\ \bibinfo
  {author} {\bibfnamefont {S.}~\bibnamefont {Walker}},\ }\href {\doibase
  10.1103/PhysRevLett.82.3316} {\bibfield  {journal} {\bibinfo  {journal}
  {Phys. Rev. Lett.}\ }\textbf {\bibinfo {volume} {82}},\ \bibinfo {pages}
  {3316} (\bibinfo {year} {1999})}\BibitemShut {NoStop}%
\bibitem [{\citenamefont {Schwartz}\ \emph {et~al.}(2011)\citenamefont
  {Schwartz}, \citenamefont {Hutchison}, \citenamefont {Genet},\ and\
  \citenamefont {Ebbesen}}]{Schwartz11}%
  \BibitemOpen
  \bibfield  {author} {\bibinfo {author} {\bibfnamefont {T.}~\bibnamefont
  {Schwartz}}, \bibinfo {author} {\bibfnamefont {J.~A.}\ \bibnamefont
  {Hutchison}}, \bibinfo {author} {\bibfnamefont {C.}~\bibnamefont {Genet}}, \
  and\ \bibinfo {author} {\bibfnamefont {T.~W.}\ \bibnamefont {Ebbesen}},\
  }\href {\doibase 10.1103/PhysRevLett.106.196405} {\bibfield  {journal}
  {\bibinfo  {journal} {Phys. Rev. Lett.}\ }\textbf {\bibinfo {volume} {106}},\
  \bibinfo {pages} {196405} (\bibinfo {year} {2011})}\BibitemShut {NoStop}%
\bibitem [{\citenamefont {Tischler}\ \emph {et~al.}(2005)\citenamefont
  {Tischler}, \citenamefont {Bradley}, \citenamefont {Bulovi\'{c}},
  \citenamefont {Song},\ and\ \citenamefont {Nurmikko}}]{Tischler2005}%
  \BibitemOpen
  \bibfield  {author} {\bibinfo {author} {\bibfnamefont {J.~R.}~\bibnamefont
  {Tischler}}, \bibinfo {author} {\bibfnamefont {M.~S.}~\bibnamefont {Bradley}},
  \bibinfo {author} {\bibfnamefont {V.}~\bibnamefont {Bulovi\'{c}}}, \bibinfo
  {author} {\bibfnamefont {J.~H.}~\bibnamefont {Song}}, \ and\ \bibinfo {author}
  {\bibfnamefont {A.}~\bibnamefont {Nurmikko}},\ }\href {\doibase
  10.1103/PhysRevLett.95.036401} {\bibfield  {journal} {\bibinfo  {journal}
  {Phys. Rev. Lett.}\ }\textbf {\bibinfo {volume} {95}},\ \bibinfo {pages}
  {036401} (\bibinfo {year} {2005})}\BibitemShut {NoStop}%
\bibitem [{\citenamefont {Dicke}(1954)}]{dicke54}%
  \BibitemOpen
  \bibfield  {author} {\bibinfo {author} {\bibfnamefont {R.}~\bibnamefont
  {Dicke}},\ }\href {\doibase 10.1103/PhysRev.93.99} {\bibfield  {journal}
  {\bibinfo  {journal} {Phys. Rev.}\ }\textbf {\bibinfo {volume} {93}},\
  \bibinfo {pages} {99} (\bibinfo {year} {1954})}\BibitemShut {NoStop}%
\bibitem [{\citenamefont {Michetti}\ \emph {et~al.}(2015)\citenamefont
  {Michetti}, \citenamefont {Mazza},\ and\ \citenamefont
  {Rocca}}]{Michetti2015}%
  \BibitemOpen
  \bibfield  {author} {\bibinfo {author} {\bibfnamefont {P.}~\bibnamefont
  {Michetti}}, \bibinfo {author} {\bibfnamefont {L.}~\bibnamefont {Mazza}}, \
  and\ \bibinfo {author} {\bibfnamefont {G.~C.~L.}\ \bibnamefont {Rocca}},\
  }\href {\doibase 10.1007/978-3-662-45082-6} {\emph {\bibinfo {title}
  {{Organic Nanophotonics}}}},\ edited by\ \bibinfo {editor} {\bibfnamefont
  {Y.~S.}\ \bibnamefont {Zhao}},\ Nano-Optics and Nanophotonics\ (\bibinfo
  {publisher} {Springer Berlin Heidelberg},\ \bibinfo {address} {Berlin,
  Heidelberg},\ \bibinfo {year} {2015})\BibitemShut {NoStop}%
\bibitem [{\citenamefont {K\'{e}na-Cohen}\ \emph {et~al.}(2008)\citenamefont
  {K\'{e}na-Cohen}, \citenamefont {Davan\c{c}o},\ and\ \citenamefont
  {Forrest}}]{Kena-Cohen2008}%
  \BibitemOpen
  \bibfield  {author} {\bibinfo {author} {\bibfnamefont {S.}~\bibnamefont
  {K\'{e}na-Cohen}}, \bibinfo {author} {\bibfnamefont {M.}~\bibnamefont
  {Davan\c{c}o}}, \ and\ \bibinfo {author} {\bibfnamefont {S.~R.}~\bibnamefont
  {Forrest}},\ }\href {\doibase 10.1103/PhysRevLett.101.116401} {\bibfield
  {journal} {\bibinfo  {journal} {Phys. Rev. Lett.}\ }\textbf {\bibinfo
  {volume} {101}},\ \bibinfo {pages} {116401} (\bibinfo {year}
  {2008})}\BibitemShut {NoStop}%
\bibitem [{\citenamefont {K\'{e}na-Cohen}\ and\ \citenamefont
  {Forrest}(2010)}]{Forrest2010}%
  \BibitemOpen
  \bibfield  {author} {\bibinfo {author} {\bibfnamefont {S.}~\bibnamefont
  {K\'{e}na-Cohen}}\ and\ \bibinfo {author} {\bibfnamefont {S.~R.}\
  \bibnamefont {Forrest}},\ }\href {\doibase 10.1038/nphoton.2010.86}
  {\bibfield  {journal} {\bibinfo  {journal} {Nat. Photon.}\ }\textbf {\bibinfo
  {volume} {4}},\ \bibinfo {pages} {371} (\bibinfo {year} {2010})}\BibitemShut
  {NoStop}%
\bibitem [{\citenamefont {Plumhof}\ \emph {et~al.}(2014)\citenamefont
  {Plumhof}, \citenamefont {St\"{o}ferle}, \citenamefont {Mai}, \citenamefont
  {Scherf},\ and\ \citenamefont {Mahrt}}]{Plumhof2013}%
  \BibitemOpen
  \bibfield  {author} {\bibinfo {author} {\bibfnamefont {J.~D.}\ \bibnamefont
  {Plumhof}}, \bibinfo {author} {\bibfnamefont {T.}~\bibnamefont
  {St\"{o}ferle}}, \bibinfo {author} {\bibfnamefont {L.}~\bibnamefont {Mai}},
  \bibinfo {author} {\bibfnamefont {U.}~\bibnamefont {Scherf}}, \ and\ \bibinfo
  {author} {\bibfnamefont {R.~F.}\ \bibnamefont {Mahrt}},\ }\href {\doibase
  10.1038/nmat3825} {\bibfield  {journal} {\bibinfo  {journal} {Nat. Mater.}\
  }\textbf {\bibinfo {volume} {13}},\ \bibinfo {pages} {247} (\bibinfo {year}
  {2014})}\BibitemShut {NoStop}%
\bibitem [{\citenamefont {Daskalakis}\ \emph {et~al.}(2014)\citenamefont
  {Daskalakis}, \citenamefont {Maier}, \citenamefont {Murray},\ and\
  \citenamefont {K\'{e}na-Cohen}}]{Daskalakis2014}%
  \BibitemOpen
  \bibfield  {author} {\bibinfo {author} {\bibfnamefont {K.~S.}\ \bibnamefont
  {Daskalakis}}, \bibinfo {author} {\bibfnamefont {S.~A.}\ \bibnamefont
  {Maier}}, \bibinfo {author} {\bibfnamefont {R.}~\bibnamefont {Murray}}, \
  and\ \bibinfo {author} {\bibfnamefont {S.}~\bibnamefont {K\'{e}na-Cohen}},\
  }\href {\doibase 10.1038/nmat3874} {\bibfield  {journal} {\bibinfo  {journal}
  {Nat. Mater.}\ }\textbf {\bibinfo {volume} {13}},\ \bibinfo {pages} {271}
  (\bibinfo {year} {2014})}\BibitemShut {NoStop}%
\bibitem [{\citenamefont {Hutchison}\ \emph {et~al.}(2012)\citenamefont
  {Hutchison}, \citenamefont {Schwartz}, \citenamefont {Genet}, \citenamefont
  {Devaux},\ and\ \citenamefont {Ebbesen}}]{hutchison12}%
  \BibitemOpen
  \bibfield  {author} {\bibinfo {author} {\bibfnamefont {J.~A.}\ \bibnamefont
  {Hutchison}}, \bibinfo {author} {\bibfnamefont {T.}~\bibnamefont {Schwartz}},
  \bibinfo {author} {\bibfnamefont {C.}~\bibnamefont {Genet}}, \bibinfo
  {author} {\bibfnamefont {E.}~\bibnamefont {Devaux}}, \ and\ \bibinfo {author}
  {\bibfnamefont {T.~W.}\ \bibnamefont {Ebbesen}},\ }\href {\doibase
  10.1002/ange.201107033} {\bibfield  {journal} {\bibinfo  {journal}
  {Angewandte Chemie}\ }\textbf {\bibinfo {volume} {124}},\ \bibinfo {pages}
  {1624} (\bibinfo {year} {2012})}\BibitemShut {NoStop}%
\bibitem [{\citenamefont {Orgiu}\ \emph {et~al.}(2015)\citenamefont {Orgiu},
  \citenamefont {George}, \citenamefont {Hutchison}, \citenamefont {Devaux},
  \citenamefont {Dayen}, \citenamefont {Doudin}, \citenamefont {Stellacci},
  \citenamefont {Genet}, \citenamefont {Samori},\ and\ \citenamefont
  {Ebbesen}}]{Orgiu2014}%
  \BibitemOpen
  \bibfield  {author} {\bibinfo {author} {\bibfnamefont {E.}~\bibnamefont
  {Orgiu}}, \bibinfo {author} {\bibfnamefont {J.}~\bibnamefont {George}},
  \bibinfo {author} {\bibfnamefont {J.~A.}\ \bibnamefont {Hutchison}}, \bibinfo
  {author} {\bibfnamefont {E.}~\bibnamefont {Devaux}}, \bibinfo {author}
  {\bibfnamefont {J.~F.}\ \bibnamefont {Dayen}}, \bibinfo {author}
  {\bibfnamefont {B.}~\bibnamefont {Doudin}}, \bibinfo {author} {\bibfnamefont
  {F.}~\bibnamefont {Stellacci}}, \bibinfo {author} {\bibfnamefont
  {C.}~\bibnamefont {Genet}}, \bibinfo {author} {\bibfnamefont
  {P.}~\bibnamefont {Samori}}, \ and\ \bibinfo {author} {\bibfnamefont {T.~W.}\
  \bibnamefont {Ebbesen}},\ }\href {\doibase
  10.1038/nmat4392} {\bibfield  {journal} {\bibinfo  {journal} {Nat. Mat.}\
  }\textbf {\bibinfo {volume} {14}},\ \bibinfo {pages} {1123} (\bibinfo {year}
  {2015})}\BibitemShut {NoStop}%
\bibitem [{\citenamefont {Schachenmayer}\ \emph {et~al.}(2014)\citenamefont
  {Schachenmayer}, \citenamefont {Genes}, \citenamefont {Tignone},\ and\
  \citenamefont {Pupillo}}]{Schachenmayer2014}%
  \BibitemOpen
  \bibfield  {author} {\bibinfo {author} {\bibfnamefont {J.}~\bibnamefont
  {Schachenmayer}}, \bibinfo {author} {\bibfnamefont {C.}~\bibnamefont
  {Genes}}, \bibinfo {author} {\bibfnamefont {E.}~\bibnamefont {Tignone}}, \
  and\ \bibinfo {author} {\bibfnamefont {G.}~\bibnamefont {Pupillo}},\ }\href {\doibase
  10.1103/PhysRevLett.114.196403} {\bibfield  {journal} {\bibinfo  {journal} {Phys. Rev. Lett.}\
  }\textbf {\bibinfo {volume} {114}},\ \bibinfo {pages} {196403} (\bibinfo {year}
  {2015})}\BibitemShut {NoStop}%
\bibitem [{\citenamefont {Feist}\ and\ \citenamefont
  {Garcia-Vidal}(2015)}]{Feist2014}%
  \BibitemOpen
  \bibfield  {author} {\bibinfo {author} {\bibfnamefont {J.}~\bibnamefont
  {Feist}}\ and\ \bibinfo {author} {\bibfnamefont {F.~J.}\ \bibnamefont
  {Garcia-Vidal}},\ }\href {\doibase
  10.1103/PhysRevLett.114.196402} {\bibfield  {journal} {\bibinfo  {journal} {Phys. Rev. Lett.}\
  }\textbf {\bibinfo {volume} {114}},\ \bibinfo {pages} {196402} (\bibinfo {year}
  {2015})}\BibitemShut {NoStop}%
\bibitem [{\citenamefont {Shalabney}\ \emph {et~al.}(2015)\citenamefont
  {Shalabney}, \citenamefont {George}, \citenamefont {Hutchison}, \citenamefont
  {Pupillo}, \citenamefont {Genet},\ and\ \citenamefont
  {Ebbesen}}]{Shalabney2014}%
  \BibitemOpen
  \bibfield  {author} {\bibinfo {author} {\bibfnamefont {A.}~\bibnamefont
  {Shalabney}}, \bibinfo {author} {\bibfnamefont {J.}~\bibnamefont {George}},
  \bibinfo {author} {\bibfnamefont {J.~A.}\ \bibnamefont {Hutchison}}, \bibinfo
  {author} {\bibfnamefont {G.}~\bibnamefont {Pupillo}}, \bibinfo {author}
  {\bibfnamefont {C.}~\bibnamefont {Genet}}, \ and\ \bibinfo {author}
  {\bibfnamefont {T.~W.}\ \bibnamefont {Ebbesen}},\ }\href {\doibase
  10.1038/ncomms6981} {\bibfield  {journal} {\bibinfo  {journal} {Nat. Comm.}\
  }\textbf {\bibinfo {volume} {6}},\ \bibinfo {pages} {5981} (\bibinfo {year}
  {2015})}\BibitemShut {NoStop}%
\bibitem [{\citenamefont {Roelli}\ \emph {et~al.}(2014)\citenamefont {Roelli},
  \citenamefont {Galland}, \citenamefont {Piro},\ and\ \citenamefont
  {Kippenberg}}]{Roelli2014}%
  \BibitemOpen
  \bibfield  {author} {\bibinfo {author} {\bibfnamefont {P.}~\bibnamefont
  {Roelli}}, \bibinfo {author} {\bibfnamefont {C.}~\bibnamefont {Galland}},
  \bibinfo {author} {\bibfnamefont {N.}~\bibnamefont {Piro}}, \ and\ \bibinfo
  {author} {\bibfnamefont {T.~J.}\ \bibnamefont {Kippenberg}},\ }\href {\doibase
  10.1038/nnano.2015.264} {\bibfield  {journal} {\bibinfo  {journal} {Nat. Nano.}\
  }\textbf {\bibinfo {volume} {advance online publication}},\ \bibinfo {pages} {} (\bibinfo {year}
  {2015})}\BibitemShut {NoStop}%
\bibitem [{\citenamefont {Pino}\ \emph {et~al.}(2015)\citenamefont {Pino},
  \citenamefont {Feist},\ and\ \citenamefont {Garcia-vidal}}]{Pino2015}%
  \BibitemOpen
  \bibfield  {author} {\bibinfo {author} {\bibfnamefont {J.}~\bibnamefont
  {Pino}}, \bibinfo {author} {\bibfnamefont {J.}~\bibnamefont {Feist}}, \ and\
  \bibinfo {author} {\bibfnamefont {F.~J.}\ \bibnamefont {Garcia-vidal}},\
  }\href {\doibase 10.1088/1367-2630/17/5/053040} {\bibfield
   {journal} {\bibinfo  {journal} {New J. Phys.}\ }\textbf
  {\bibinfo {volume} {17}},\ \bibinfo {pages} {053040} (\bibinfo {year}
  {2015})}\BibitemShut {NoStop}%
\bibitem [{\citenamefont {Pino}\ \emph {et~al.}(2015)\citenamefont {Pino},
  \citenamefont {Feist},\ and\ \citenamefont {Garcia-vidal}}]{Pino2015b}%
  \BibitemOpen
  \bibfield  {author} {\bibinfo {author} {\bibfnamefont {J.}~\bibnamefont
  {Pino}}, \bibinfo {author} {\bibfnamefont {J.}~\bibnamefont {Feist}}, \ and\
  \bibinfo {author} {\bibfnamefont {F.~J.}\ \bibnamefont {Garcia-vidal}},\
  }\href@noop {} {\enquote {\bibinfo {title} {{Signatures of Vibrational Strong Coupling in Raman Scattering}},}\ }
  (\bibinfo {year} {2015}),\ \Eprint {http://arxiv.org/abs/1511.02115v1}
  {arXiv:1511.02115v1} \BibitemShut {NoStop}%
\bibitem [{\citenamefont {Spano}\ (2015)\citenamefont
  {Spano}}]{Spano2015}%
  \BibitemOpen
  \bibfield  {author} {\bibinfo {author} {\bibfnamefont {F.~C.}~\bibnamefont
  {Spano}},\ }\href {\doibase 10.1063/1.4919348} {\bibfield
   {journal} {\bibinfo  {journal} {J. Chem. Phys.}\ }\textbf
  {\bibinfo {volume} {142}},\ \bibinfo {pages} {184707} (\bibinfo {year}
  {2015})}\BibitemShut {NoStop}%
\bibitem [{\citenamefont {Canaguier-Durand}\ \emph {et~al.}(2013)\citenamefont
  {Canaguier-Durand}, \citenamefont {Devaux}, \citenamefont {George},
  \citenamefont {Pang}, \citenamefont {Hutchison}, \citenamefont {Schwartz},
  \citenamefont {Genet}, \citenamefont {Wilhelms}, \citenamefont {Lehn},\ and\
  \citenamefont {Ebbesen}}]{Canaguier-Durand2013}%
  \BibitemOpen
  \bibfield  {author} {\bibinfo {author} {\bibfnamefont {A.}~\bibnamefont
  {Canaguier-Durand}}, \bibinfo {author} {\bibfnamefont {E.}~\bibnamefont
  {Devaux}}, \bibinfo {author} {\bibfnamefont {J.}~\bibnamefont {George}},
  \bibinfo {author} {\bibfnamefont {Y.}~\bibnamefont {Pang}}, \bibinfo {author}
  {\bibfnamefont {J.~A.}\ \bibnamefont {Hutchison}}, \bibinfo {author}
  {\bibfnamefont {T.}~\bibnamefont {Schwartz}}, \bibinfo {author}
  {\bibfnamefont {C.}~\bibnamefont {Genet}}, \bibinfo {author} {\bibfnamefont
  {N.}~\bibnamefont {Wilhelms}}, \bibinfo {author} {\bibfnamefont {J.-M.}\
  \bibnamefont {Lehn}}, \ and\ \bibinfo {author} {\bibfnamefont {T.~W.}\
  \bibnamefont {Ebbesen}},\ }\href {\doibase 10.1002/anie.201301861} {\bibfield
   {journal} {\bibinfo  {journal} {Angew. Chem. Int. Ed. Engl.}\ }\textbf
  {\bibinfo {volume} {52}},\ \bibinfo {pages} {10533} (\bibinfo {year}
  {2013})}\BibitemShut {NoStop}%
\bibitem [{\citenamefont {Schwartz}\ \emph {et~al.}(2013)\citenamefont
  {Schwartz}, \citenamefont {Hutchison}, \citenamefont {L\'{e}onard},
  \citenamefont {Genet}, \citenamefont {Haacke},\ and\ \citenamefont
  {Ebbesen}}]{Schwartz2013a}%
  \BibitemOpen
  \bibfield  {author} {\bibinfo {author} {\bibfnamefont {T.}~\bibnamefont
  {Schwartz}}, \bibinfo {author} {\bibfnamefont {J.~A.~}\ \bibnamefont
  {Hutchison}}, \bibinfo {author} {\bibfnamefont {J.}~\bibnamefont
  {L\'{e}onard}}, \bibinfo {author} {\bibfnamefont {C.}~\bibnamefont {Genet}},
  \bibinfo {author} {\bibfnamefont {S.}~\bibnamefont {Haacke}}, \ and\ \bibinfo
  {author} {\bibfnamefont {T.~W.}\ \bibnamefont {Ebbesen}},\ }\href {\doibase
  10.1002/cphc.201200734} {\bibfield  {journal} {\bibinfo  {journal} {Chem.
  Phys. Chem}\ }\textbf {\bibinfo {volume} {14}},\ \bibinfo {pages} {125}
  (\bibinfo {year} {2013})}\BibitemShut {NoStop}%
\bibitem [{\citenamefont {Houdr\'{e}}\ \emph {et~al.}(1996)\citenamefont
  {Houdr\'{e}}, \citenamefont {Stanley},\ and\ \citenamefont
  {Ilegems}}]{houdre96}%
  \BibitemOpen
  \bibfield  {author} {\bibinfo {author} {\bibfnamefont {R.}~\bibnamefont
  {Houdr\'{e}}}, \bibinfo {author} {\bibfnamefont {R.~P.}\ \bibnamefont
  {Stanley}}, \ and\ \bibinfo {author} {\bibfnamefont {M.}~\bibnamefont
  {Ilegems}},\ }\href {\doibase 10.1103/PhysRevA.53.2711} {\bibfield  {journal}
  {\bibinfo  {journal} {Phys. Rev. A}\ }\textbf {\bibinfo {volume} {53}},\
  \bibinfo {pages} {2711} (\bibinfo {year} {1996})}\BibitemShut {NoStop}%
\bibitem [{\citenamefont {Eastham}\ and\ \citenamefont
  {Littlewood}(2001)}]{Eastham2001}%
  \BibitemOpen
  \bibfield  {author} {\bibinfo {author} {\bibfnamefont {P.~R.}\ \bibnamefont
  {Eastham}}\ and\ \bibinfo {author} {\bibfnamefont {P.~B.}\ \bibnamefont
  {Littlewood}},\ }\href {\doibase 10.1103/PhysRevB.64.235101} {\bibfield
  {journal} {\bibinfo  {journal} {Phys. Rev. B}\ }\textbf {\bibinfo {volume}
  {64}},\ \bibinfo {pages} {235101} (\bibinfo {year} {2001})}\BibitemShut
  {NoStop}%
\bibitem [{\citenamefont {Keeling}\ \emph {et~al.}(2005)\citenamefont
  {Keeling}, \citenamefont {Eastham}, \citenamefont {Szymanska},\ and\
  \citenamefont {Littlewood}}]{JKeeling2005}%
  \BibitemOpen
  \bibfield  {author} {\bibinfo {author} {\bibfnamefont {J.}~\bibnamefont
  {Keeling}}, \bibinfo {author} {\bibfnamefont {P.~R.}~\bibnamefont {Eastham}},
  \bibinfo {author} {\bibfnamefont {M.~H.}~\bibnamefont {Szymanska}}, \ and\
  \bibinfo {author} {\bibfnamefont {P.~B.}~\bibnamefont {Littlewood}},\ }\href
  {\doibase 10.1103/PhysRevB.72.115320} {\bibfield  {journal} {\bibinfo
  {journal} {Phys. Rev. B}\ }\textbf {\bibinfo {volume} {72}},\ \bibinfo
  {pages} {115320} (\bibinfo {year} {2005})}\BibitemShut {NoStop}%
  \bibitem [{\citenamefont {Ciuti}\ \emph {et~al.}(2005)\citenamefont {Ciuti},
  \citenamefont {Bastard},\ and\ \citenamefont {Carusotto}}]{Ciuti05}%
  \BibitemOpen 
  \bibfield  {author} {\bibinfo {author} {\bibfnamefont {C.}~\bibnamefont
  {Ciuti}}, \bibinfo {author} {\bibfnamefont {G.}~\bibnamefont {Bastard}}, \
  and\ \bibinfo {author} {\bibfnamefont {I.}~\bibnamefont {Carusotto}},\ }\href
  {\doibase 10.1103/PhysRevB.72.115303} {\bibfield  {journal} {\bibinfo
  {journal} {Phys. Rev. B}\ }\textbf {\bibinfo {volume} {72}},\ \bibinfo
  {pages} {115303} (\bibinfo {year} {2005})}\BibitemShut {NoStop}%
\bibitem [{\citenamefont {De~Liberato}(2015)}]{DeLiberato15}%
  \BibitemOpen
  \bibfield  {author} {\bibinfo {author} {\bibfnamefont {S.}~\bibnamefont
  {De~Liberato}},\ }\href {\doibase 10.1103/PhysRevB.92.125433} {\bibfield
  {journal} {\bibinfo  {journal} {Phys. Rev. B}\ }\textbf {\bibinfo
  {volume} {92}},\ \bibinfo {pages} {125433} (\bibinfo {year}
  {2015})}\BibitemShut {NoStop}%
\bibitem [{\citenamefont {Anappara}\ \emph {et~al.}(2009)\citenamefont
  {Anappara}, \citenamefont {De~Liberato}, \citenamefont {Tredicucci},
  \citenamefont {Ciuti}, \citenamefont {Biasiol}, \citenamefont {Sorba},\ and\
  \citenamefont {Beltram}}]{Anappara09}%
  \BibitemOpen
  \bibfield  {author} {\bibinfo {author} {\bibfnamefont {A.~A.}\ \bibnamefont
  {Anappara}}, \bibinfo {author} {\bibfnamefont {S.}~\bibnamefont
  {De~Liberato}}, \bibinfo {author} {\bibfnamefont {A.}~\bibnamefont
  {Tredicucci}}, \bibinfo {author} {\bibfnamefont {C.}~\bibnamefont {Ciuti}},
  \bibinfo {author} {\bibfnamefont {G.}~\bibnamefont {Biasiol}}, \bibinfo
  {author} {\bibfnamefont {L.}~\bibnamefont {Sorba}}, \ and\ \bibinfo {author}
  {\bibfnamefont {F.}~\bibnamefont {Beltram}},\ }\href {\doibase
  10.1103/PhysRevB.79.201303} {\bibfield  {journal} {\bibinfo  {journal} {Phys.
  Rev. B}\ }\textbf {\bibinfo {volume} {79}},\ \bibinfo {pages} {201303}
  (\bibinfo {year} {2009})}\BibitemShut {NoStop}%
\bibitem [{\citenamefont {Gambino}\ \emph {et~al.}(2014)\citenamefont
  {Gambino}, \citenamefont {Mazzeo}, \citenamefont {Genco}, \citenamefont
  {Di~Stefano}, \citenamefont {Savasta}, \citenamefont {Patanè}, \citenamefont
  {Ballarini}, \citenamefont {Mangione}, \citenamefont {Lerario}, \citenamefont
  {Sanvitto},\ and\ \citenamefont {Gigli}}]{Gambino14}%
  \BibitemOpen
  \bibfield  {author} {\bibinfo {author} {\bibfnamefont {S.}~\bibnamefont
  {Gambino}}, \bibinfo {author} {\bibfnamefont {M.}~\bibnamefont {Mazzeo}},
  \bibinfo {author} {\bibfnamefont {A.}~\bibnamefont {Genco}}, \bibinfo
  {author} {\bibfnamefont {O.}~\bibnamefont {Di~Stefano}}, \bibinfo {author}
  {\bibfnamefont {S.}~\bibnamefont {Savasta}}, \bibinfo {author} {\bibfnamefont
  {S.}~\bibnamefont {Patanè}}, \bibinfo {author} {\bibfnamefont
  {D.}~\bibnamefont {Ballarini}}, \bibinfo {author} {\bibfnamefont
  {F.}~\bibnamefont {Mangione}}, \bibinfo {author} {\bibfnamefont
  {G.}~\bibnamefont {Lerario}}, \bibinfo {author} {\bibfnamefont
  {D.}~\bibnamefont {Sanvitto}}, \ and\ \bibinfo {author} {\bibfnamefont
  {G.}~\bibnamefont {Gigli}},\ }\href {\doibase 10.1021/ph500266d} {\bibfield
  {journal} {\bibinfo  {journal} {ACS Photonics}\ }\textbf {\bibinfo {volume}
  {1}},\ \bibinfo {pages} {1042} (\bibinfo {year} {2014})}\BibitemShut
  {NoStop}%
\bibitem [{\citenamefont {Gubbin}\ \emph {et~al.}(2014)\citenamefont {Gubbin},
  \citenamefont {Maier},\ and\ \citenamefont {Kena-Cohen}}]{Gubbin14}%
  \BibitemOpen
  \bibfield  {author} {\bibinfo {author} {\bibfnamefont {C.}~\bibnamefont
  {Gubbin}}, \bibinfo {author} {\bibfnamefont {S.}~\bibnamefont {Maier}}, \
  and\ \bibinfo {author} {\bibfnamefont {S.}~\bibnamefont {Kena-Cohen}},\
  }\href@noop {} {\bibfield  {journal} {\bibinfo  {journal} {Appl. Phys.
  Lett.}\ }\textbf {\bibinfo {volume} {104}} (\bibinfo {year}
  {2014})}\BibitemShut {NoStop}%
\bibitem [{\citenamefont {Maissen}\ \emph {et~al.}(2014)\citenamefont
  {Maissen}, \citenamefont {Scalari}, \citenamefont {Valmorra}, \citenamefont
  {Beck}, \citenamefont {Faist}, \citenamefont {Cibella}, \citenamefont
  {Leoni}, \citenamefont {Reichl}, \citenamefont {Charpentier},\ and\
  \citenamefont {Wegscheider}}]{Maissen14}%
  \BibitemOpen
  \bibfield  {author} {\bibinfo {author} {\bibfnamefont {C.}~\bibnamefont
  {Maissen}}, \bibinfo {author} {\bibfnamefont {G.}~\bibnamefont {Scalari}},
  \bibinfo {author} {\bibfnamefont {F.}~\bibnamefont {Valmorra}}, \bibinfo
  {author} {\bibfnamefont {M.}~\bibnamefont {Beck}}, \bibinfo {author}
  {\bibfnamefont {J.}~\bibnamefont {Faist}}, \bibinfo {author} {\bibfnamefont
  {S.}~\bibnamefont {Cibella}}, \bibinfo {author} {\bibfnamefont
  {R.}~\bibnamefont {Leoni}}, \bibinfo {author} {\bibfnamefont
  {C.}~\bibnamefont {Reichl}}, \bibinfo {author} {\bibfnamefont
  {C.}~\bibnamefont {Charpentier}}, \ and\ \bibinfo {author} {\bibfnamefont
  {W.}~\bibnamefont {Wegscheider}},\ }\href {\doibase
  10.1103/PhysRevB.90.205309} {\bibfield  {journal} {\bibinfo  {journal} {Phys.
  Rev. B}\ }\textbf {\bibinfo {volume} {90}},\ \bibinfo {pages} {205309}
  (\bibinfo {year} {2014})}\BibitemShut {NoStop}%
\bibitem [{\citenamefont {Rzazewski}\ \emph {et~al.}(1975)\citenamefont
  {Rzazewski}, \citenamefont {W\'{o}dkiewicz},\ and\ \citenamefont
  {Zakowicz}}]{Rzazewski1975}%
  \BibitemOpen
  \bibfield  {author} {\bibinfo {author} {\bibfnamefont {K.}~\bibnamefont
  {Rzazewski}}, \bibinfo {author} {\bibfnamefont {K.}~\bibnamefont
  {W\'{o}dkiewicz}}, \ and\ \bibinfo {author} {\bibfnamefont {W.}~\bibnamefont
  {Zakowicz}},\ }\href {\doibase 10.1103/PhysRevLett.35.432} {\bibfield
  {journal} {\bibinfo  {journal} {Phys. Rev. Lett.}\ }\textbf {\bibinfo
  {volume} {35}},\ \bibinfo {pages} {432} (\bibinfo {year} {1975})}\BibitemShut
  {NoStop}%
\bibitem [{\citenamefont {Ciuti}\ and\ \citenamefont
  {Carusotto}(2006)}]{Ciuti2006}%
  \BibitemOpen
  \bibfield  {author} {\bibinfo {author} {\bibfnamefont {C.}~\bibnamefont
  {Ciuti}}\ and\ \bibinfo {author} {\bibfnamefont {I.}~\bibnamefont
  {Carusotto}},\ }\href {\doibase 10.1103/PhysRevA.74.033811} {\bibfield
  {journal} {\bibinfo  {journal} {Phys. Rev. A}\ }\textbf {\bibinfo {volume}
  {74}},\ \bibinfo {pages} {033811} (\bibinfo {year} {2006})}\BibitemShut
  {NoStop}%
\bibitem [{\citenamefont {Schrieffer}\ and\ \citenamefont
  {Wolff}(1966)}]{schrieffer66}%
  \BibitemOpen
  \bibfield  {author} {\bibinfo {author} {\bibfnamefont {J.~R.}\ \bibnamefont
  {Schrieffer}}\ and\ \bibinfo {author} {\bibfnamefont {P.~A.}\ \bibnamefont
  {Wolff}},\ }\href {\doibase 10.1103/PhysRev.149.491} {\bibfield  {journal}
  {\bibinfo  {journal} {Phys. Rev.}\ }\textbf {\bibinfo {volume} {149}},\
  \bibinfo {pages} {491} (\bibinfo {year} {1966})}\BibitemShut {NoStop}%
\bibitem [{Note1()}]{Note1}%
  \BibitemOpen
\bibinfo {note} {Note however that the collective splitting $g_\protect
  \mathbf {k}\protect \sqrt {N_m}$ can however still be comparable to $\omega
  _k+\epsilon $, as is indeed the case for the parameters we
  consider.}\BibitemShut {Stop}%
  \bibitem [{\citenamefont {Casanova}\ \emph {et~al.}(2010)\citenamefont
  {Casanova}, \citenamefont {Romero}, \citenamefont {Lizuain}, \citenamefont
  {Garcia-Ripoll},\ and\ \citenamefont {Solano}}]{Casanova10}%
  \BibitemOpen
  \bibfield  {author} {\bibinfo {author} {\bibfnamefont {J.}~\bibnamefont
  {Casanova}}, \bibinfo {author} {\bibfnamefont {G.}~\bibnamefont {Romero}},
  \bibinfo {author} {\bibfnamefont {I.}~\bibnamefont {Lizuain}}, \bibinfo
  {author} {\bibfnamefont {J.~J.}\ \bibnamefont {Garcia-Ripoll}}, \ and\
  \bibinfo {author} {\bibfnamefont {E.}~\bibnamefont {Solano}},\ }\href
  {\doibase 10.1103/PhysRevLett.105.263603} {\bibfield  {journal} {\bibinfo
  {journal} {Phys. Rev. Lett.}\ }\textbf {\bibinfo {volume} {105}},\ \bibinfo
  {pages} {263603} (\bibinfo {year} {2010})}\BibitemShut {NoStop}%
\bibitem [{\citenamefont {De~Liberato}(2014)}]{DeLiberato14}%
  \BibitemOpen
  \bibfield  {author} {\bibinfo {author} {\bibfnamefont {S.}~\bibnamefont
  {De~Liberato}},\ }\href {\doibase 10.1103/PhysRevLett.112.016401} {\bibfield
  {journal} {\bibinfo  {journal} {Phys. Rev. Lett.}\ }\textbf {\bibinfo
  {volume} {112}},\ \bibinfo {pages} {016401} (\bibinfo {year}
  {2014})}\BibitemShut {NoStop}%
\bibitem [{\citenamefont {Kim}\ \emph {et~al.}(2015)\citenamefont
  {Kim}, \citenamefont {Sim}, \citenamefont {Yoon}, \citenamefont {Gong}, \citenamefont {Ahn}, \citenamefont {Cho},\ and\ \citenamefont {Lee}}]{Kim15}%
  \BibitemOpen
  \bibfield  {author} {\bibinfo {author} {\bibfnamefont {M.-K.}~\bibnamefont
  {Kim}}, \bibinfo {author} {\bibfnamefont {H.}~\bibnamefont {Sim}},
  \bibinfo {author} {\bibfnamefont {S.~J.}~\bibnamefont {Yoon}}, \bibinfo {author} {\bibfnamefont {S.-H.}~\bibnamefont {Gong}}, \bibinfo {author} {\bibfnamefont {C.~W.}~\bibnamefont {Ahn}}, \bibinfo {author} {\bibfnamefont {Y.-H.}~\bibnamefont {Cho}},\ and\
  \bibinfo {author} {\bibfnamefont {Y.-H.}~\bibnamefont {Lee}},\ }\href
  {\doibase 10.1021/acs.nanolett.5b01204} {\bibfield  {journal} {\bibinfo
  {journal} {Nano Lett.}\ }\textbf {\bibinfo {volume} {15}},\ \bibinfo
  {pages} {4120} (\bibinfo {year} {2015})}\BibitemShut {NoStop}%
\bibitem [{\citenamefont {Caldwell}\ \emph {et~al.}(2013)\citenamefont
  {Caldwell}, \citenamefont {Glembockl}, \citenamefont {Francescato}, \citenamefont {Sharac}, \citenamefont {Glannini}, \citenamefont {Bezares}, \citenamefont {Long}, \citenamefont {Owrutsky}, \citenamefont {Vurgaftman}, \citenamefont {Tischler}, \citenamefont {Wheeler}, \citenamefont {Bassim}, \citenamefont {Shirey}, \citenamefont {Kaslca},\ and\ \citenamefont {Maler}}]{Caldwell13}%
  \BibitemOpen
  \bibfield  {author} {\bibinfo {author} {\bibfnamefont {J.~D.}~\bibnamefont
  {Caldwell}}, \bibinfo {author} {\bibfnamefont {O.~J.}~\bibnamefont {Glembockl}},
  \bibinfo {author} {\bibfnamefont {Y.}~\bibnamefont {Francescato}}, \bibinfo {author} {\bibfnamefont {N.}~\bibnamefont {Sharac}}, \bibinfo {author} {\bibfnamefont {V.}~\bibnamefont {Glannini}}, \bibinfo {author} {\bibfnamefont {F.~J.}~\bibnamefont {Bezares}},  \bibinfo {author} {\bibfnamefont {J.~P.}~\bibnamefont {Long}},  \bibinfo {author} {\bibfnamefont {J.~C.}~\bibnamefont {Owrutsky}},  \bibinfo {author} {\bibfnamefont {I.}~\bibnamefont {Vurgaftman}},  \bibinfo {author} {\bibfnamefont {J.~G.}~\bibnamefont {Tischler}},  \bibinfo {author} {\bibfnamefont {V.~D.}~\bibnamefont {Wheeler}}, \bibinfo {author} {\bibfnamefont {N.~D.}~\bibnamefont {Bassim}}, \bibinfo {author} {\bibfnamefont {L.~M.}~\bibnamefont {Shirey}}, \bibinfo {author} {\bibfnamefont {R.}~\bibnamefont {Kaslca}},\ and\
  \bibinfo {author} {\bibfnamefont {S.-A.}~\bibnamefont {Maler}},\ }\href
  {\doibase 10.1021/nl401590g} {\bibfield  {journal} {\bibinfo
  {journal} {Nano Lett.}\ }\textbf {\bibinfo {volume} {13}},\ \bibinfo
  {pages} {3690} (\bibinfo {year} {2013})}\BibitemShut {NoStop}%
\bibitem [{\citenamefont {De~Liberato}\ \emph {et~al.}(2009)\citenamefont
  {De~Liberato}, \citenamefont {Gerace}, \citenamefont {Carusotto},\ and\ \citenamefont {Ciuti}}]{DeLiberato09}%
  \BibitemOpen
  \bibfield  {author} {\bibinfo {author} {\bibfnamefont {S.}~\bibnamefont
  {De~Liberato}}, \bibinfo {author} {\bibfnamefont {D.}~\bibnamefont {Gerace}},
  \bibinfo {author} {\bibfnamefont {I.}~\bibnamefont {Carusotto}}, \ and\
  \bibinfo {author} {\bibfnamefont {C.}~\bibnamefont {Ciuti}},\ }\href
  {\doibase 10.1103/PhysRevA.80.053810} {\bibfield  {journal} {\bibinfo
  {journal} {Phys. Rev. A}\ }\textbf {\bibinfo {volume} {80}},\ \bibinfo
  {pages} {053810} (\bibinfo {year} {2009})}\BibitemShut {NoStop}%
\bibitem [{Note2()}]{Note2}%
  \BibitemOpen
  \bibinfo {note} {One may note that if the vibrational modes were degenerate
  $\Omega _1=\Omega _2=\Omega $, then the effective Huang-Rhys parameter will
  be $S_{\protect \text {eff}}=S_1+S_2$}\BibitemShut {NoStop}%
\bibitem [{\citenamefont {T\"orm\"a}\ and\ \citenamefont
  {Barnes}(2015)}]{torma15}%
  \BibitemOpen
  \bibfield  {author} {\bibinfo {author} {\bibfnamefont {P.}~\bibnamefont
  {T\"orm\"a}}\ and\ \bibinfo {author} {\bibfnamefont {W.~L.}\ \bibnamefont
  {Barnes}},\ }\href {http://stacks.iop.org/0034-4885/78/i=1/a=013901}
  {\bibfield  {journal} {\bibinfo  {journal} {Reports on Progress in Physics}\
  }\textbf {\bibinfo {volume} {78}},\ \bibinfo {pages} {013901} (\bibinfo
  {year} {2015})}\BibitemShut {NoStop}%
\bibitem [{\citenamefont {Carusotto}\ and\ \citenamefont
  {Ciuti}(2013)}]{Carusotto2013}%
  \BibitemOpen
  \bibfield  {author} {\bibinfo {author} {\bibfnamefont {I.}~\bibnamefont
  {Carusotto}}\ and\ \bibinfo {author} {\bibfnamefont {C.}~\bibnamefont
  {Ciuti}},\ }\href {\doibase 10.1103/RevModPhys.85.299} {\bibfield  {journal}
  {\bibinfo  {journal} {Rev. Mod. Phys.}\ }\textbf {\bibinfo {volume} {85}},\
  \bibinfo {pages} {299} (\bibinfo {year} {2013})}\BibitemShut {NoStop}%
\bibitem [{\citenamefont {Kasprzak}\ \emph {et~al.}(2006)\citenamefont
  {Kasprzak}, \citenamefont {Richard}, \citenamefont {Kundermann},
  \citenamefont {Baas}, \citenamefont {Jeambrun}, \citenamefont {Keeling},
  \citenamefont {Marchetti}, \citenamefont {Szymanska}, \citenamefont
  {Andr\'{e}}, \citenamefont {Staehli}, \citenamefont {Savona}, \citenamefont
  {Littlewood}, \citenamefont {Deveaud},\ and\ \citenamefont
  {Dang}}]{Kasprzak2006}%
  \BibitemOpen
  \bibfield  {author} {\bibinfo {author} {\bibfnamefont {J.}~\bibnamefont
  {Kasprzak}}, \bibinfo {author} {\bibfnamefont {M.}~\bibnamefont {Richard}},
  \bibinfo {author} {\bibfnamefont {S.}~\bibnamefont {Kundermann}}, \bibinfo
  {author} {\bibfnamefont {A.}~\bibnamefont {Baas}}, \bibinfo {author}
  {\bibfnamefont {P.}~\bibnamefont {Jeambrun}}, \bibinfo {author}
  {\bibfnamefont {J.~M.~J.}\ \bibnamefont {Keeling}}, \bibinfo {author}
  {\bibfnamefont {F.~M.}\ \bibnamefont {Marchetti}}, \bibinfo {author}
  {\bibfnamefont {M.~H.}\ \bibnamefont {Szymanska}}, \bibinfo {author}
  {\bibfnamefont {R.}~\bibnamefont {Andr\'{e}}}, \bibinfo {author}
  {\bibfnamefont {J.~L.}\ \bibnamefont {Staehli}}, \bibinfo {author}
  {\bibfnamefont {V.}~\bibnamefont {Savona}}, \bibinfo {author} {\bibfnamefont
  {P.~B.}\ \bibnamefont {Littlewood}}, \bibinfo {author} {\bibfnamefont
  {B.}~\bibnamefont {Deveaud}}, \ and\ \bibinfo {author} {\bibfnamefont
  {L.~S.}\ \bibnamefont {Dang}},\ }\href {\doibase 10.1038/nature05131}
  {\bibfield  {journal} {\bibinfo  {journal} {Nature}\ }\textbf {\bibinfo
  {volume} {443}},\ \bibinfo {pages} {409} (\bibinfo {year}
  {2006})}\BibitemShut {NoStop}%
\bibitem [{\citenamefont {Balili}\ \emph {et~al.}(2007)\citenamefont {Balili},
  \citenamefont {Hartwell}, \citenamefont {Snoke}, \citenamefont {Pfeiffer},\
  and\ \citenamefont {West}}]{snoke07science}%
  \BibitemOpen
  \bibfield  {author} {\bibinfo {author} {\bibfnamefont {R.}~\bibnamefont
  {Balili}}, \bibinfo {author} {\bibfnamefont {V.}~\bibnamefont {Hartwell}},
  \bibinfo {author} {\bibfnamefont {D.}~\bibnamefont {Snoke}}, \bibinfo
  {author} {\bibfnamefont {L.}~\bibnamefont {Pfeiffer}}, \ and\ \bibinfo
  {author} {\bibfnamefont {K.}~\bibnamefont {West}},\ }\href {\doibase
  10.1126/science.1140990} {\bibfield  {journal} {\bibinfo  {journal} {Science
  (80-. ).}\ }\textbf {\bibinfo {volume} {316}},\ \bibinfo {pages} {1007}
  (\bibinfo {year} {2007})}\BibitemShut {NoStop}%
\bibitem [{\citenamefont {Klaers}\ \emph {et~al.}(2010)\citenamefont {Klaers},
  \citenamefont {Schmitt}, \citenamefont {Vewinger},\ and\ \citenamefont
  {Weitz}}]{Klaers2010}%
  \BibitemOpen
  \bibfield  {author} {\bibinfo {author} {\bibfnamefont {J.}~\bibnamefont
  {Klaers}}, \bibinfo {author} {\bibfnamefont {J.}~\bibnamefont {Schmitt}},
  \bibinfo {author} {\bibfnamefont {F.}~\bibnamefont {Vewinger}}, \ and\
  \bibinfo {author} {\bibfnamefont {M.}~\bibnamefont {Weitz}},\ }\href
  {\doibase 10.1038/nature09567} {\bibfield  {journal} {\bibinfo  {journal}
  {Nature}\ }\textbf {\bibinfo {volume} {468}},\ \bibinfo {pages} {545}
  (\bibinfo {year} {2010})}\BibitemShut {NoStop}%
\bibitem [{\citenamefont {Litinskaya}\ \emph {et~al.}(2004)\citenamefont
  {Litinskaya}, \citenamefont {Reineker},\ and\ \citenamefont
  {Agranovich}}]{Litinskaya2004}%
  \BibitemOpen
  \bibfield  {author} {\bibinfo {author} {\bibfnamefont {M.}~\bibnamefont
  {Litinskaya}}, \bibinfo {author} {\bibfnamefont {P.}~\bibnamefont
  {Reineker}}, \ and\ \bibinfo {author} {\bibfnamefont {V.~M.}\ \bibnamefont
  {Agranovich}},\ }\href {\doibase 10.1002/pssa.200304067} {\bibfield
  {journal} {\bibinfo  {journal} {Phys. Stat. Sol.}\ }\textbf {\bibinfo
  {volume} {201}},\ \bibinfo {pages} {646} (\bibinfo {year}
  {2004})}\BibitemShut {NoStop}%
\bibitem [{\citenamefont {Litinskaya}\ and\ \citenamefont
  {Reineker}(2006)}]{Litinskaya2006}%
  \BibitemOpen
  \bibfield  {author} {\bibinfo {author} {\bibfnamefont {M.}~\bibnamefont
  {Litinskaya}}\ and\ \bibinfo {author} {\bibfnamefont {P.}~\bibnamefont
  {Reineker}},\ }\href {\doibase 10.1103/PhysRevB.74.165320} {\bibfield
  {journal} {\bibinfo  {journal} {Phys. Rev. B}\ }\textbf {\bibinfo {volume}
  {74}},\ \bibinfo {pages} {165320} (\bibinfo {year} {2006})}\BibitemShut
  {NoStop}%
\bibitem [{\citenamefont {Litinskaya}(2008)}]{Litinskaya2008}%
  \BibitemOpen
  \bibfield  {author} {\bibinfo {author} {\bibfnamefont {M.}~\bibnamefont
  {Litinskaya}},\ }\href {\doibase 10.1103/PhysRevB.77.155325} {\bibfield
  {journal} {\bibinfo  {journal} {Phys. Rev. B}\ }\textbf {\bibinfo {volume}
  {77}},\ \bibinfo {pages} {155325} (\bibinfo {year} {2008})}\BibitemShut
  {NoStop}%
\bibitem [{\citenamefont {Michetti}\ and\ \citenamefont {{La
  Rocca}}(2009)}]{Michetti2009}%
  \BibitemOpen
  \bibfield  {author} {\bibinfo {author} {\bibfnamefont {P.}~\bibnamefont
  {Michetti}}\ and\ \bibinfo {author} {\bibfnamefont {G.~C.}~\bibnamefont {{La
  Rocca}}},\ }\href {\doibase 10.1103/PhysRevB.79.035325} {\bibfield  {journal}
  {\bibinfo  {journal} {Phys. Rev. B}\ }\textbf {\bibinfo {volume} {79}},\
  \bibinfo {pages} {035325} (\bibinfo {year} {2009})}\BibitemShut {NoStop}%
\bibitem [{\citenamefont {Fontanesi}\ and\ \citenamefont {{La
  Rocca}}(2009)}]{Fontanesi2009}%
  \BibitemOpen
  \bibfield  {author} {\bibinfo {author} {\bibfnamefont {L.}~\bibnamefont
  {Fontanesi}},\ {\bibfnamefont {L.}~\bibnamefont
  {Mazza}}, and\ \bibinfo {author} {\bibfnamefont {G.~C.}\ \bibnamefont
  {{La Rocca}}},\ }\href {\doibase 10.1103/PhysRevB.80.235313} {\bibfield
  {journal} {\bibinfo  {journal} {Phys. Rev. B}\ }\textbf {\bibinfo {volume}
  {80}},\ \bibinfo {pages} {235313} (\bibinfo {year} {2009})}\BibitemShut
  {NoStop}%
\bibitem [{\citenamefont {Mazza}\ and\ \citenamefont {{La
  Rocca}}(2009)}]{Mazza2009}%
  \BibitemOpen
  \bibfield  {author} {\bibinfo {author} {\bibfnamefont {L.}~\bibnamefont
  {Mazza}},\ {\bibfnamefont {L.}~\bibnamefont
  {Fontanesi}} and\ \bibinfo {author} {\bibfnamefont {G.~C.}\ \bibnamefont {{La
  Rocca}}},\ }\href {\doibase 10.1103/PhysRevB.80.235314} {\bibfield  {journal}
  {\bibinfo  {journal} {Phys. Rev. B}\ }\textbf {\bibinfo {volume} {80}},\
  \bibinfo {pages} {235314} (\bibinfo {year} {2009})}\BibitemShut {NoStop}%
\bibitem [{\citenamefont {Mazza}\ \emph {et~al.}(2013)\citenamefont {Mazza},
  \citenamefont {K\'{e}na-Cohen}, \citenamefont {Michetti},\ and\ \citenamefont
  {{La Rocca}}}]{Mazza2013}%
  \BibitemOpen
  \bibfield  {author} {\bibinfo {author} {\bibfnamefont {L.}~\bibnamefont
  {Mazza}}, \bibinfo {author} {\bibfnamefont {S.}~\bibnamefont
  {K\'{e}na-Cohen}}, \bibinfo {author} {\bibfnamefont {P.}~\bibnamefont
  {Michetti}}, \ and\ \bibinfo {author} {\bibfnamefont {G.~C.}\ \bibnamefont
  {{La Rocca}}},\ }\href {\doibase 10.1103/PhysRevB.88.075321} {\bibfield
  {journal} {\bibinfo  {journal} {Phys. Rev. B}\ }\textbf {\bibinfo {volume}
  {88}},\ \bibinfo {pages} {075321} (\bibinfo {year} {2013})}\BibitemShut
  {NoStop}%
\bibitem [{\citenamefont {\'{C}wik}\ \emph {et~al.}(2014)\citenamefont
  {\'{C}wik}, \citenamefont {Reja}, \citenamefont {Littlewood},\ and\
  \citenamefont {Keeling}}]{Cwik2014}%
  \BibitemOpen
  \bibfield  {author} {\bibinfo {author} {\bibfnamefont {J.~A.}\ \bibnamefont
  {\'{C}wik}}, \bibinfo {author} {\bibfnamefont {S.}~\bibnamefont {Reja}},
  \bibinfo {author} {\bibfnamefont {P.~B.}\ \bibnamefont {Littlewood}}, \ and\
  \bibinfo {author} {\bibfnamefont {J.}~\bibnamefont {Keeling}},\ }\href
  {\doibase 10.1209/0295-5075/105/47009} {\bibfield  {journal} {\bibinfo
  {journal} {Eur. Lett.}\ }\textbf {\bibinfo {volume} {105}},\ \bibinfo {pages}
  {47009} (\bibinfo {year} {2014})}\BibitemShut {NoStop}%
\bibitem [{\citenamefont {Galego}\ \emph {et~al.}(2015)\citenamefont {Galego},
  \citenamefont {Garcia-Vidal},\ and\ \citenamefont {Feist}}]{Galego2015}%
  \BibitemOpen
  \bibfield  {author} {\bibinfo {author} {\bibfnamefont {J.}~\bibnamefont
  {Galego}}, \bibinfo {author} {\bibfnamefont {F.~J.}\ \bibnamefont
  {Garcia-Vidal}}, \ and\ \bibinfo {author} {\bibfnamefont {J.}~\bibnamefont
  {Feist}},\ }\href {\doibase 10.1103/PhysRevX.5.041022} {\bibfield
  {journal} {\bibinfo  {journal} {Phys. Rev. X}\ }\textbf {\bibinfo {volume}
  {5}},\ \bibinfo {pages} {041022} (\bibinfo {year} {2015})}\BibitemShut
  {NoStop}%
\bibitem [{\citenamefont {Abrikosov}\ \emph {et~al.}(1975)\citenamefont
  {Abrikosov}, \citenamefont {Gorkov},\ and\ \citenamefont
  {Dzyaloshinski}}]{agd}%
  \BibitemOpen
  \bibfield  {author} {\bibinfo {author} {\bibfnamefont {A.}~\bibnamefont
  {Abrikosov}}, \bibinfo {author} {\bibfnamefont {L.}~\bibnamefont {Gorkov}}, \
  and\ \bibinfo {author} {\bibfnamefont {I.}~\bibnamefont {Dzyaloshinski}},\
  }\href@noop {} {\emph {\bibinfo {title} {{Methods of Quantum Field Theory in
  Statistical Physics}}}}\ (\bibinfo  {publisher} {Dover},\ \bibinfo {address}
  {New York},\ \bibinfo {year} {1975})\BibitemShut {NoStop}%
\bibitem [{\citenamefont {Collett}\ and\ \citenamefont
  {Gardiner}(1984)}]{Collett1984}%
  \BibitemOpen
  \bibfield  {author} {\bibinfo {author} {\bibfnamefont {M.~J.}\ \bibnamefont
  {Collett}}\ and\ \bibinfo {author} {\bibfnamefont {C.~W.}\ \bibnamefont
  {Gardiner}},\ }\href {\doibase 10.1103/PhysRevA.30.1386} {\bibfield
  {journal} {\bibinfo  {journal} {Phys. Rev. A}\ }\textbf {\bibinfo {volume}
  {30}},\ \bibinfo {pages} {1386} (\bibinfo {year} {1984})}\BibitemShut
  {NoStop}%
\end{thebibliography}
%

\end{document}